\documentclass[sigconf]{acmart} %review, anonymous (this option removed) 

\usepackage{subfigure}
\usepackage{subcaption}

\usepackage{amsmath,amsfonts}
\usepackage[algo2e,linesnumbered,ruled,lined,noend]{algorithm2e}
\SetKwInOut{Parameter}{Parameters}
\SetKwInOut{Constraint}{Constraints}
\SetKwFunction{Mining}{Mining}
\SetKwFunction{MiningTwo}{Mining2}
\SetKwFunction{Candidates}{Candidates}
\SetKwProg{Fn}{Function}{}{}
\usepackage{graphicx}
\usepackage{textcomp}
\usepackage{color}
\usepackage{xcolor}
\usepackage{dsfont}
\usepackage{bbm}
\usepackage{placeins} % preamble

\newcommand{\std}[1]{{\scriptsize (#1)}}
\usepackage[dvipsnames]{xcolor}

\usepackage{xspace}
\def\BibTeX{{\rm B\kern-.05em{\sc i\kern-.025em b}\kern-.08em
    T\kern-.1667em\lower.7ex\hbox{E}\kern-.125emX}}
    
\usepackage{multirow} % Required for multirows
\usepackage{booktabs} % For prettier tables
\usepackage{colortbl}

\usepackage[normalem]{ulem}
\useunder{\uline}{\ul}{}

\usepackage{pgf}
\usepackage{enumitem}
\setlist[itemize]{leftmargin=*}
\usepackage{stfloats}

\usepackage{makecell}

\usepackage{bbold}
\usepackage{mathtools}
\usepackage{comment}
\usepackage{microtype}
\usepackage[misc]{ifsym}
\usepackage{url}
\usepackage{xspace}

\SetCommentSty{mycommfont}

\newcommand{\smallsection}[1]{{\vspace{0.02in} \noindent {{\underline{\smash{\bf #1:}}}}}}
\newcommand{\method}{\textsc{PerPEFT}\xspace}

\definecolor{sunwoogreen}{RGB}{32, 200, 150}
\definecolor{sunwoogreen2}{RGB}{67, 148, 58}
\definecolor{sunwooyellow}{rgb}{1.0, 1.0, 0.0}
\definecolor{sunwooyellow2}{RGB}{228, 208, 10}

\newcommand{\best}{\cellcolor{sunwoogreen!70}}  %{0.9}
\newcommand{\gframe}[1]{\fcolorbox{white}{sunwoogreen!70}{\strut #1}}

\AtBeginDocument{%
  \providecommand\BibTeX{{%
    \normalfont B\kern-0.5em{\scshape i\kern-0.25em b}\kern-0.8em\TeX}}}

\copyrightyear{2026}
\acmYear{2026}
\setcopyright{cc}
\setcctype{by}
\acmConference[WWW '26]{Proceedings of the ACM Web Conference 2026}{April 13--17, 2026}{Dubai, United Arab Emirates}
\acmBooktitle{Proceedings of the ACM Web Conference 2026 (WWW '26), April 13--17, 2026, Dubai, United Arab Emirates}
\acmPrice{}
\acmDOI{10.1145/3774904.3792248}
\acmISBN{979-8-4007-2307-0/2026/04}
\usepackage{needspace}



\begin{document}

    \title[Personalized Parameter-Efficient Fine-Tuning of Foundation Models for Multimodal Recommendation]{Personalized Parameter-Efficient Fine-Tuning of \\ Foundation Models for Multimodal Recommendation}
    
    \settopmatter{authorsperrow=3}
	
    \author{Sunwoo Kim}
	\affiliation{%
    	\institution{KAIST}
            \city{Seoul}
            \country{Republic of Korea}
	}
	\email{kswoo97@kaist.ac.kr}

    \author{Hyunjin Hwang}
	\affiliation{%
		\institution{KAIST}
            \city{Seoul}
            \country{Republic of Korea}
	}
	\email{hyunjinhwang@kaist.ac.kr}

	\author{Kijung Shin}
	\affiliation{%
		\institution{KAIST}
            \city{Seoul}
            \country{Republic of Korea}
	}
	\email{kijungs@kaist.ac.kr}
	
    \begin{abstract}

In recent years, substantial research has integrated multimodal item metadata into recommender systems, often by using pre‑trained multimodal foundation models to encode such data. 
Since these models are not originally trained for recommendation tasks, recent works efficiently adapt them via parameter‑efficient fine‑tuning (PEFT). 
{However, even with PEFT, item embeddings from multimodal foundation models remain user-blind}: item embeddings are not conditioned on user interests, despite the fact that users with diverse interests attend to different item aspects.
To address this limitation, we propose \method, a personalized PEFT strategy for multimodal recommendation. 
Specifically, \method groups users by interest and assigns a distinct PEFT module to each group, enabling each module to capture the fine‑grained item aspects most predictive of that group’s purchase decisions. 
We further introduce a specialized training technique that strengthens this user‑group conditioning. 
Notably, \method is PEFT‑agnostic and can be paired with any PEFT method applicable to multimodal foundation models. 
Through extensive experiments, we show that (1) \method outperforms the strongest baseline by {up to 15.3\% (NDCG@20)} and (2) delivers consistent gains across diverse PEFT variants.
It is noteworthy that, even with personalization, PEFT remains lightweight, adding only 1.3\% of the parameter count of the foundation model.
We provide our code and datasets at ~\url{https://github.com/kswoo97/PerPEFT}.

    \end{abstract}
	
	\maketitle

    \section{Introduction}
    \label{sec:intro}

Recommender systems play a central role in modern web services, responsible for retrieving items that closely align with user interests from extensive item pools~\cite{liu2025modalsync, wang2025unleashing, kim2024towards, kim2025itemrag}. 
By inferring user interests from historical purchase behavior, they provide personalized recommendations that both enhance user satisfaction and drive business revenue.

In recent years, substantial efforts have been devoted to integrating \textit{multimodal item metadata} into recommender systems~\cite{liu2024multimodal, zhao2025dvib, attimonelli2024ducho, wei2024promptmm, wang2025generative, zhang2025hierarchical}.
Such data often includes textual information (e.g., item titles and descriptions) and visual information (e.g., item images).

\begin{figure}[t] 
  \centering
  \includegraphics[width=1.0\linewidth]{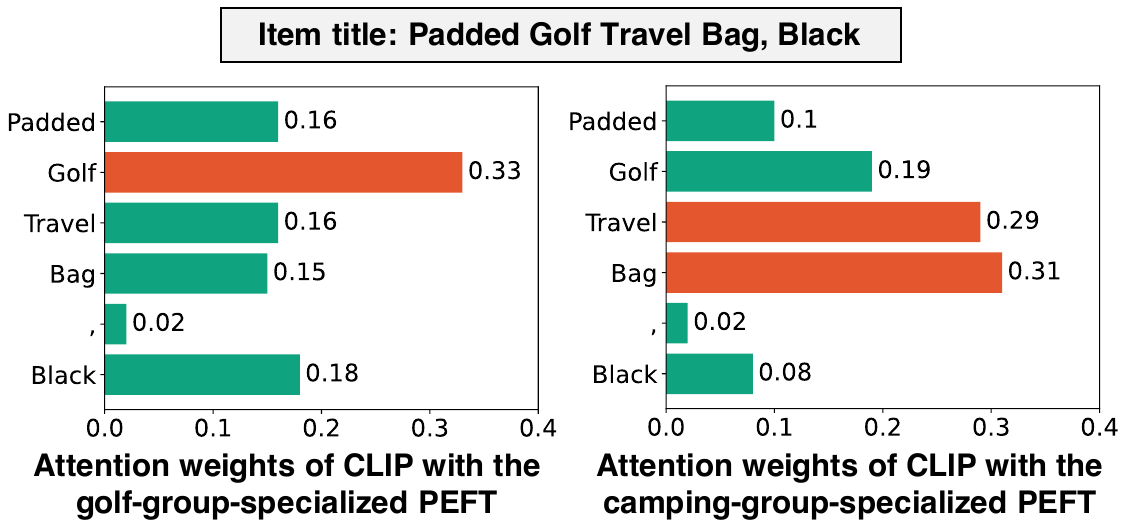} % or .png/.jpg
  \caption{\textit{\method guides CLIP~\cite{radford2021learning}, a multimodal foundation model, to focus on item aspects aligned with each user group's interests.}
  For a golf travel bag, CLIP {personalized to a golf-interest group} attends to `Golf', while CLIP z{personalized to} a camping-interest group focuses on `Travel' and `Bag'.}
  \label{fig:mainattention}
\end{figure}

One promising line of research transforms multimodal data into vector representations (i.e., embeddings) using pre-trained multimodal foundation models (e.g., CLIP~\cite{radford2021learning})~\cite{liu2025joint, xu2025cohesion, pomo2025recommender}.
The representations serve either as core item representations~\cite{yuan2023go, xu2025mdvt} or as supplementary components within recommender systems~\cite{zhang2021mining, zhou2023tale}.

Among them, a notable branch explores fine-tuning multimodal foundation models~\cite{yuan2023go, yi2025enhancing}.
Since such models are not pre-trained for recommendation tasks, fine-tuning the models for this purpose enhances their recommendation performance~\cite{yuan2023go}, enabling them to better capture item characteristics particularly useful for inferring user preferences, such as type and brand.

Given the immense cost of foundation-model fine-tuning, recent work has adopted \textit{parameter-efficient fine-tuning} (PEFT) strategies for multimodal recommendation~\cite{fu2024iisan}.
PEFT trains only a small set of additional parameters, while keeping the majority of the pre-trained model parameters frozen~\cite{han2024peftsurvey, fu2024exploring}.
In multimodal recommendation tasks, PEFT substantially reduces the time and memory required for fine-tuning, while still achieving performance comparable to full fine-tuning, which updates all parameters of foundation models~\cite{fu2024iisan}.

Beyond simply applying PEFT, a key open question is \textit{what desirable properties should multimodal foundation models exhibit in multimodal recommendation?}
Since these models often act as item encoders that capture aspects of items influencing purchases, we focus on how users consider these aspects in their purchase decisions.
In particular, users with diverse interests naturally emphasize different aspects of items.
For example, character-figure collectors tend to focus on appearance-related aspects (e.g., colorway and design), whereas game enthusiasts often consider factors such as genre (e.g., FPS or RPG) and platform (e.g., PC or console).
Therefore, it is desirable for multimodal foundation models to capture different item aspects in accordance with user interests.

In this work, we introduce \method (\textbf{\underline{Per}}sonalized \textbf{\underline{P}}arameter-\textbf{\underline{E}}fficient \textbf{\underline{F}}ine-\textbf{\underline{T}}uning for multimodal recommendation), a PEFT strategy that allows multimodal foundation models to focus on different item characteristics depending on users’ distinct interests.
Notably, \method is PEFT-agnostic and can be coupled with any PEFT module applicable to multimodal foundation models.
In a nutshell, \method groups users with similar interests and assigns a distinct PEFT module to each user group, with each module trained separately on its corresponding users.
Through this separation, each module is guided to focus on the item aspects that users with the corresponding interests are likely to value.
We further enhance \method with a specialized training strategy that uses group-specific hard negative sampling to improve its capacity to learn purchase-relevant item aspects distinctive to each user group.

\begin{example}[Foundation Models Personalized by \method]
{Figure~\ref{fig:mainattention} shows that \method personalizes CLIP~\cite{radford2021learning}, a multimodal foundation model, to emphasize item aspects reflecting each user group’s interests.
For a golf travel bag, CLIP personalized to a golf-interest group places its highest attention on the `Golf' token.
In contrast, CLIP personalized to a camping-interest group assigns relatively higher attention to `Travel' and `Bag' tokens. 
Further details are provided in Section~\ref{subsec:designobjective}.}
\end{example}

Through extensive experiments on four real-world e-commerce benchmark datasets against seven baseline methods, we demonstrate the effectiveness of \method for multimodal recommendation. 
Specifically, \method (1) outperforms existing personalization techniques in 44 out of 48 settings, (2) remains consistently effective when integrated with diverse PEFT modules, and (3) focuses on different aspects of items according to users' varying interests.
Notably, even with our personalization, the parameter overhead is minimal: the PEFT modules add only 1.3\% as many parameters as CLIP, our multimodal foundation model.

Our key contributions are summarized as follows.

\begin{itemize}
    \item \textbf{C1) New concept.} We introduce the concept of personalized PEFT and propose \method to align foundation models with users’ diverse interests for multimodal recommendation.
    \item \textbf{C2) Training strategy.} We also propose an effective training strategy for \method that enables each PEFT module to capture fine-grained item attributes specific to the associated user group.
    \item \textbf{C3) Empirical validation.} We experimentally demonstrate the superiority of \method over existing baseline methods, achieving up to 15.3\% performance gain on NDCG@20 in the Arts, Crafts \& Sewing dataset, compared to the strongest baseline.
\end{itemize}
For \textbf{reproducibility}, we provide our code and datasets at ~\url{https://github.com/kswoo97/PerPEFT}.

% Despite its effectiveness, fine-tuning all parameters of multimodal foundation models is often expensive and may even be infeasible under limited computational resources. 
% This difficulty comes from their massive scale, with state-of-the-art models containing nearly a billion parameters~\cite{li2024multimodal}.

%% 앞쪽 빌드업 부분을 지금보다 좀 더 간소화

% To address this issue, recent work has applied \textit{parameter-efficient fine-tuning} (PEFT) strategies in multimodal recommendation~\cite{fu2024iisan}.

    \section{Related work and preliminary}
    \label{sec:relatedwork}

In this section, we review related studies and present the preliminary concepts relevant to our work.

\subsection{Related work}

We discuss two research branches relevant to our study. 
% multimodal recommendation and parameter-efficient fine-tuning (PEFT).

\subsubsection{\textbf{Multimodal recommendation}}
Beyond user–item interactions, many studies have incorporated multimodal item metadata into recommender systems~\cite{chen2025hypercomplex, pomo2025recommender}. % ; refer to surveys~\cite{liu2024multimodal, liu2024multimodal2}.
Recent approaches employ item representations that are obtained from their text and image with a pre-trained multimodal foundation model~\cite{zhou2023tale, zhang2021mining}.
Among these, a prominent direction fine-tunes multimodal foundation models for the recommendation task~\cite{zhang2024multimodal, fu2024efficient, yuan2023go}.
They attach multimodal representations to a backbone recommendation model (e.g., SASRec~\cite{kang2018self}) and fine-tune the foundation model by propagating gradients from the recommendation loss.
{While improving recommendation accuracy, these methods} incur significant computational costs by updating foundation models with many parameters~\cite{fu2024iisan}.
To alleviate this, PEFT has been incorporated for multimodal recommendation~\cite{fu2024iisan}. % , as discussed below

\subsubsection{\textbf{Parameter-efficient fine-tuning (PEFT)}}\label{subsec:peft}
Significant efforts have focused on efficiently fine-tuning pre-trained foundation models to improve their performance on targeted downstream tasks~\cite{han2024peftsurvey, liu2024moe, ma2022scattered, fu2024exploring}.
Among various PEFT methods, LoRA~\cite{hu2021lora} and (IA)$^{3}$~\cite{liu2022fewshot} stand out.
LoRA reparameterizes each weight update of a pre-trained model into two smaller low-rank matrices, training only these matrices instead of directly updating the full weight parameters.
In contrast, (IA)$^{3}$ introduces lightweight gates that rescale intermediate embeddings within the foundation model on a dimension-by-dimension basis, enabling selective emphasis of informative dimensions.
\citet{fu2024iisan} proposed IISAN, a PEFT method specially designed for multimodal recommendation.
By attaching adapters externally rather than embedding them within backbone layers, IISAN avoids expensive backpropagation within the foundation model, making it substantially faster during training. 

% side-network-based 

% as general techniques for Transformers~\cite{vaswani2017attention}, the backbone architecture of most multimodal foundation models~\cite{radford2021learning,jia2021scaling, li2022blip}

%IISAN achieves notable training speedups, but it does not take user interests into account in the multimodal encoding process.

% \noindent{\underline{\textbf{Note.}}} To the best of our knowledge, existing PEFT methods for multimodal recommendation have not incorporated personalization into the multimodal encoding process, yielding the same multimodal embedding for an item regardless of the target user.

\subsection{Preliminary}\label{subsec:preliminary}
The user set and the item set are denoted by $\mathcal{U}=\{u_{1},u_{2},\cdots u_{\vert \mathcal{U}\vert }\}$ and $\mathcal{I}=\{i_{1},i_{2},\cdots,i_{\vert \mathcal{I}\vert}\}$, respectively.
We consider a sequential recommendation setting, and therefore, each user $u_{t} \in \mathcal{U}$ is represented by her purchase history sequence.\footnote{{We present results under non-sequential recommender systems in Appendix~\ref{app:lightgcn}.}}
Each item $i_{k} \in \mathcal{I}$ is associated with an image ${x}^{(img)}_{k}$ and a textual description $y^{(txt)}_{k}$, in addition to a learnable transductive embedding $\mathbf{h}_{k}\in\mathbb{R}^{d}$.
We consider a multimodal foundation model $f$ that receives an image-text pair $(x^{(img)}_{k},y^{(txt)}_{k})$ and returns the corresponding visual and textual embeddings $\mathbf{x}_{k}\in\mathbb{R}^{d'}$ and $\mathbf{y}_{k} \in \mathbb{R}^{d'}$ (i.e., $f_{\theta}(x^{(img)}_{k},y^{(txt)}_{k}) = (\mathbf{x}_{k}, \mathbf{y}_{k})$).
A PEFT module $\theta$ is attached to the foundation model $f$, with the resulting foundation model denoted as $f_{\theta}$.
Note that during training, the pre-trained parameters of $f$ remain frozen, while only the parameters of the PEFT module $\theta$ are updated for the recommendation task.
Further details regarding PEFT methods are in Appendix~\ref{app:peftdetail}.

    \section{Proposed method}
    \label{sec:method}

In this section, we introduce \method (\textbf{\underline{Per}}sonalized \textbf{\underline{P}}arameter-\textbf{\underline{E}}fficient \textbf{\underline{F}}ine-\textbf{\underline{T}}uning for multimodal recommendation), a novel PEFT strategy that equips multimodal foundation models with the ability to derive distinct item embeddings conditioned on user interests.
We begin by describing \textit{Global PEFT}, our baseline backbone PEFT method (Section~\ref{subsec:methodbaseline}).
We then introduce the overall pipeline of our PEFT approach, \method (Section~\ref{subsec:methodencoding}).
Lastly, we elaborate on our specialized training technique for \method (Section~\ref{subsec:methodtraining}) and {analyze its parameter overhead (Section~\ref{subsec:paramoverhead}).}

\begin{figure*}[t] 
    \centering
    \includegraphics[width=0.9\linewidth]{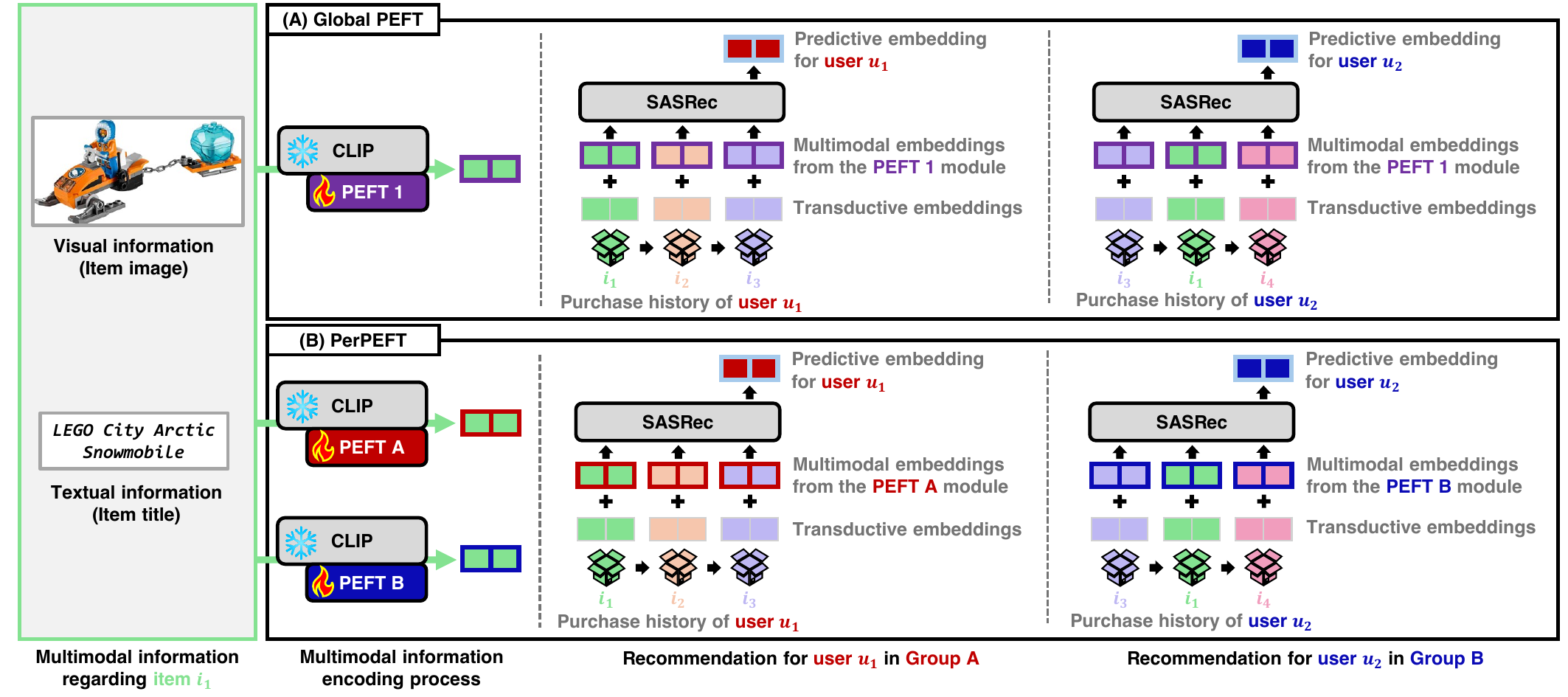} 
    \vspace{-1mm}
    \caption{\textit{An example case of (A) Global PEFT and (B) \method, our personalized PEFT for multimodal recommendation}. 
    Each item’s multimodal information is encoded by CLIP, a multimodal foundation model, with an attached PEFT module.
    Unlike Global PEFT, which uses the same PEFT module for users $u_{1}$ and $u_{2}$, \method employs different PEFT modules according to {their inferred interest groups}. % each user’s group.
    Subsequently, we form each item embedding by summing the multimodal and transductive embeddings. 
    To generate recommendations for a user, we construct the sequence of item embeddings in the same order as the user's purchase history and feed it into SASRec, the backbone recommender system.}
    \label{fig:mainfigure}
    \vspace{-1mm}
\end{figure*}

\subsection{Naive approach: Global PEFT}\label{subsec:methodbaseline}
Global PEFT is our baseline method, which shares a single PEFT module across all users.
We first describe how Global PEFT derives item embeddings from multimodal information, and then describe how recommendation is performed using these embeddings.
An example pipeline is illustrated in the block (A) of Figure~\ref{fig:mainfigure}.

\subsubsection{\textbf{Item encoding}}
We use CLIP~\cite{radford2021learning} as the multimodal foundation model $f$.
Its PEFT-attached version is denoted as $f_{\theta}$, where $\theta$ is the attached PEFT module, such as LoRA~\cite{hu2021lora} or IISAN~\cite{fu2024iisan}.
% We denote the multimodal foundation model $f$, which is CLIP~\cite{radford2021learning} in our work, with an attached PEFT module $\theta$ as $f_{\theta}$.
To compute the embedding of item $i_{k}$, denoted by $\mathbf{z}_{k} \in \mathbb{R}^{d}$, we first obtain its multimodal embedding $\mathbf{m}_{k} \in \mathbb{R}^{d}$ with the PEFT-attached CLIP $f_{\theta}$.
Specifically, we obtain the visual and textual embeddings $\mathbf{x}_{k} \in \mathbb{R}^{d'}$ and $\mathbf{y}_{k} \in \mathbb{R}^{d'}$, from its image ${x}^{(img)}_{k}$ and textual description $y^{(txt)}_{k}$ using $f_{\theta}$: $f_{\theta}(x^{(img)}_{k},y^{(txt)}_{k}) = (\mathbf{x}_{k}, \mathbf{y}_{k})$.
These embeddings are concatenated and projected through a projector MLP to form the multimodal embedding: $\texttt{MLP}([\mathbf{x}_{k}\vert\vert\mathbf{y}_{k}])=\mathbf{m}_{k}$, where $[\mathbf{x} \vert\vert \mathbf{y}]$ denotes the concatenation of vectors $\mathbf{x}$ and $\mathbf{y}$.
Finally, we add $\mathbf{m}_{k}$ with the item’s transductive embedding $\mathbf{h}_{k}$ to form the final item embedding: $\mathbf{m}_{k} + \mathbf{h}_{k} = \mathbf{z}_{k}$.

\subsubsection{\textbf{Recommendation process}}\label{subsec:globaltraining}
The obtained item embeddings are then fed into the backbone recommender system, which in our work is {SASRec}~\cite{kang2018self}, a representative sequential recommender system. 
We consider recommending items to user $u_{t}$ with purchase history $[i_{t,1}, i_{t,2}, \cdots ,i_{t,(n-1)}, i_{t,n}]$; the associated embeddings are $\mathbf{z}_{t,1},\mathbf{z}_{t,2}, \cdots ,\mathbf{z}_{t,(n-1)}, \mathbf{z}_{t,n} \in \mathbb{R}^{d}$.
% \footnote{While we use {SASRec} as our backbone recommender system, other representative models (e.g., Bert4Rec~\cite{sun2019bert4rec} and S$^{3}$-Rec~\cite{zhou2020s3}) are also applicable.}

During training, the embeddings without the last item's embedding are fed into SASRec, and it returns the predictive embeddings $\mathbf{z}' \in \mathbb{R}^{d}$ for each input item: $\texttt{SASRec}([\mathbf{z}_{t,1},\mathbf{z}_{t,2},\cdots ,\mathbf{z}_{t,(n-1)}]) = [\mathbf{z}'_{t,1}, \mathbf{z}'_{t,2},\cdots \mathbf{z}'_{t,(n-1)}]$.
Then, each embedding is trained to yield high recommendation scores for its corresponding next item, where each score is defined as the inner product with the next item's embedding {(i.e., $\mathbf{z}_{t,(p+1)}^T\mathbf{z}'_{t,p}$, for each $p \in \{1,2,\cdots ,(n-1)\}$)}.
To this end, following common {SASRec} training pipeline~\cite{kang2018self}, we use the binary cross-entropy loss with negative sampling:
% \begin{equation}\label{eq:globalloss}
%      \log{(\sigma(\mathbf{z}^{T}_{a}\mathbf{z}'_{k}))} +\log{(\sigma(\mathbf{z}^{T}_{b}\mathbf{z}'_{p}))} - \log{(\sigma(\mathbf{z}^{T}_{p}\mathbf{z}'_{k}))} - \log{(\sigma(\mathbf{z}^{T}_{q}\mathbf{z}'_{p}))},
% \end{equation}
\begin{equation}\label{eq:globalloss}
    \mathcal{L}^{(glob)}_{t}=\sum_{p=1}^{n-1}\log{\sigma(\mathbf{z}^{T}_{a_{p}}\mathbf{z}'_{t,p})} -\sum^{n-1}_{p=1}\log{\sigma(\mathbf{z}^{T}_{t,(p+1)}\mathbf{z}'_{t,p})}
\end{equation}
where $\{i_{a_{p}}: \forall p\in \{1,2,\cdots , (n-1) \}\} \subset \mathcal{I}$ are negative samples from $\mathcal{I}$, and $\sigma(\cdot)$ is a sigmoid function.
Lastly, the parameters of (1) PEFT module $\theta$, (2) transductive item embeddings $\mathbf{h}$, (3) {MLP} projector, and (4) backbone SASRec are updated to minimize $\mathcal{L}_{t}^{(glob)}$. 
%(Eq.~\eqref{eq:globalloss}) via gradient descent.

During final prediction, the full purchase sequence of user $u_{t}$ (i.e., $[\mathbf{z}_{t,1},\mathbf{z}_{t,2}, \cdots , \mathbf{z}_{t,n}]$) is fed into {SASRec} and it generates predictive embeddings as: $\texttt{SASRec}([\mathbf{z}_{t,1},\mathbf{z}_{t,2}, \cdots , \mathbf{z}_{t,n}]) = [\mathbf{z}'_{t,1},\mathbf{z}'_{t,2}, \cdots , \mathbf{z}'_{t,n}]$.
The predictive embedding of the last item, $\mathbf{z}'_{t,n}$, is used as the final predictive embedding for user $u{_t}$, denoted as $\mathbf{u}_{t}$ (i.e., $\mathbf{u}_{t} \coloneqq \mathbf{z}'_{t,n}$). 
The recommendation score for each item is then computed as $ \mathbf{u}_{t}^{\top}\mathbf{z}_{j} \in \mathbb{R}, \ \forall i_{j} \in \mathcal{I}$, and the Top-$K$ items with the highest scores are recommended to the user.

\subsection{Proposed method: \method}\label{subsec:methodencoding}
We now introduce \method, a personalized PEFT method for multimodal recommendation.
We first give an overview of \method, and then elaborate on how \method is designed. 
An example pipeline is illustrated in the block (B) of Figure~\ref{fig:mainfigure}.

\subsubsection{\textbf{High-level idea.}}\label{subsec:ourhighlevel}
The central aim of \method is to enable the multimodal foundation model to focus on different aspects across different user interest domains {(see Figure~\ref{fig:mainattention} for an example)}, allowing it to generate item representations conditioned on user interests.
To achieve this, \method groups users with similar interests and assigns distinct PEFT modules to each group. 
For example, in the block (B) of Figure~\ref{fig:mainfigure}, user $u_{1}$ and user $u_{2}$ belong to different groups; thus, different PEFT modules are applied for them. 
Therefore, the same item is encoded into different multimodal embeddings depending on the user.
Then, each PEFT module is trained exclusively on the users belonging to its assigned group. 
As a result, different user groups provide disjoint training samples, encouraging each module to specialize in its corresponding group.

% \subsubsection{\textbf{Advantages of grouping.}}
% Alternatively, distinct PEFT modules can be assigned to individual users for more fine-grained personalization. 
% However, although PEFT modules are much lighter than multimodal foundation models, they still involve a non-negligible number of parameters. 
% Therefore, training such modules on the purchase history of a single user often leads to the risk of overfitting due to the limited number of training data.
% We corroborate this by demonstrating that recommendation performance declines once the number of user groups exceeds a certain threshold, as detailed in Appendix~\red{X}.
% Thus, in light of this practical challenge, we find that user grouping is a more effective strategy for achieving personalization in PEFT for multimodal recommendation.

\subsubsection{\textbf{User grouping}}\label{subsec:grouping}
To group users according to their interests, we begin by obtaining their interest representations.
{To this end, we employ the predictive embedding $\mathbf{u}_{t}$ of each user $u_t$} derived by Global PEFT, as detailed in Section~\ref{subsec:methodbaseline}.
Specifically, we train (1) PEFT modules, (2) transductive item embeddings, (3) MLP projector, and (4) backbone SASRec.
After training, we compute a predictive embedding $\mathbf{u}_{t}$ for each user $u_{t} \in \mathcal{U}$ as detailed in Section~\ref{subsec:globaltraining}, and use it as the user's interest representation. 
Our rationale is that since these embeddings are optimized to predict the next item the user will purchase, they effectively capture the user's preferences.
Using the obtained representations, we apply K-Means clustering~\cite{arthur2007k}, and treat each resulting cluster as a user group with similar interests.
{Note that our approach is not limited to K-Means and exhibits comparable performance with other clustering algorithms (see Appendix~\ref{app:otherclustering}).}
Formally, we denote each disjoint user group as $\mathcal{U}^{(c)}, \forall c \in \{1,2,\cdots ,C\}$, such that $\mathcal{U} = \bigcup_{c=1}^{C} \mathcal{U}^{(c)}$.

\subsubsection{\textbf{Multimodal information encoding.}}\label{subsec:persencoding}
In \method, each user group is assigned a distinct PEFT module and MLP projector to generate multimodal item embeddings. 
For instance, when obtaining the representation of item $i_{k}$ for a user in $\mathcal{U}^{(c)}$, we employ the PEFT module and MLP projector specific to that group, denoted by $\theta^{(c)}$ and \texttt{MLP}$^{(c)}$, respectively.
\method adopts the same encoding framework as Global PEFT (Section~\ref{subsec:methodbaseline}), with the key distinction that the components are group-specific ($\theta^{(c)}$ and \texttt{MLP}$^{(c)}$), producing item representations specialized for each group.
We denote the multimodal embedding of item $i_{k}$ specialized for $\mathcal{U}^{(c)}$ as $\mathbf{m}^{(c)}_{k}$.
% Note that, {despite the group-specific components, the increase in the number of parameters in practice is not significant 
% (see Section~\ref{subsec:paramoverhead}).}

\subsubsection{\textbf{Final item embeddings and recommendation process}}\label{subsec:finalrecommendation}
While the multimodal information encoding process is specialized for each user group, we share the transductive item embeddings and the backbone recommender system (SASRec) across all groups.
Our choice is motivated by two factors: (1) assigning separate transductive embeddings and backbone models would introduce substantially more parameters, and (2) performance remains competitive even when these components are shared across all groups, as studied in our preliminary analysis.
Thus, except for the use of group-specific multimodal embeddings, \method performs the recommendation task in the same way as Global PEFT.

\begin{figure}[t] 
    \centering
    \includegraphics[width=0.9\linewidth]{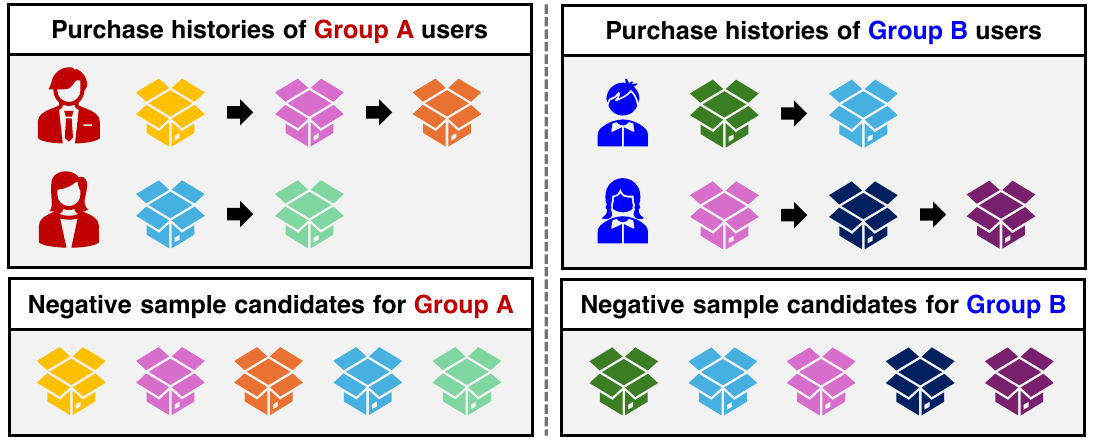} 
    \caption{\textit{An example case of our training technique for \method}. 
    When training for Group A, we draw negative samples only from items appearing in Group A users’ purchase histories, instead of the full item set.
    The same holds for Group B.
    }
    \label{fig:trainingfigure}
\end{figure}

\subsection{Training technique for \method}\label{subsec:methodtraining}

In this section, we introduce our specialized training technique for \method, which uses \textit{group-specific negative sampling}.
We begin by outlining the challenge in \method and proceed to our training remedy.
An example case of our technique is provided in Figure~\ref{fig:trainingfigure}.

\subsubsection{\textbf{Challenge.}}
Recall that \method groups users by similar interests, attaches a distinct PEFT module to each group, and trains each module on that group’s users (Section~\ref{subsec:ourhighlevel}).
Except for this, \method adopts the Global PEFT pipeline; accordingly, we optimize all learnable parameters using binary cross-entropy with negative sampling (Eq.~\eqref{eq:globalloss}; Section~\ref{subsec:globaltraining}). 
% As briefly discussed in Section~\ref{subsec:finalrecommendation}, all learnable parameters within \method are trained through the binary cross-entropy loss with the negative sampling, as in Global PEFT (Eq.~\eqref{eq:globalloss}; Section~\ref{subsec:globaltraining}). 
However, while Global PEFT obtains negative samples from the full item set $\mathcal{I}$, applying the same protocol to \method does not improve recommendation performance compared to Global PEFT (see ablation, Section~\ref{subsec:ablation}).
We hypothesize that the underperformance stems from a mismatch between positive item space and negative sample space. 
Specifically, each group’s positives are the items its users actually purchased, while negatives are drawn from the full item set $\mathcal{I}$, where many of them never appear as positives. 
As a result, the PEFT module and other components may easily reduce the loss without actually learning the item's group-specific aspects, which are beneficial for accurate recommendations.

\subsubsection{\textbf{Remedy.}}\label{subsec:remedy}
To mitigate this issue, we employ group-specific negative sampling, where negatives are selected from each group’s item pool (i.e., items purchased at least once by users in that group), and the corresponding PEFT module is updated accordingly.
This approach yields harder negative samples that encourage the PEFT module to capture more accurate, purchase-relevant item aspects. 
This intuition aligns with prior findings in general representation learning~\cite{robinson2021contrastive, radenovic2023filtering, patel2024tripletclip}, which report that hard negative samples can enhance representation quality.

\subsubsection{\textbf{Training detail.}}\label{subsec:ourdetail}
Before training, we initialize (1) the group-specific components (PEFT modules $\theta^{(c)}$ and projector \texttt{MLP}$^{(c)}, \forall c \in \{1,2,\cdots,C\}$) and (2) the global components (transductive item embeddings and SASRec) with the corresponding parameters from the model trained via Global PEFT (Section~\ref{subsec:methodbaseline}), which is also being used for user grouping (Section~\ref{subsec:grouping}).
Then, let $\mathcal{I}^{(c)} \subseteq \mathcal{I}$ denote a set of items purchased at least once by users in $\mathcal{U}^{(c)}$.
Consider user $u_{t}$ belonging $\mathcal{U}^{(c)}$, with purchase history $[i_{t,1}, i_{t,2}, \cdots ,i_{t,(n-1)}, i_{t,n}]$.
Given multimodal embeddings $\mathbf{m}^{(c)}_{t,1},\mathbf{m}^{(c)}_{t,2},\cdots ,\mathbf{m}^{(c)}_{t,(n-1)},\mathbf{m}^{(c)}_{t,n} \in \mathbb{R}^{d}$ obtained as in Section~\ref{subsec:persencoding}, the final embeddings are $\mathbf{z}^{(c)}_{t,p}=\mathbf{h}_{t,p}+\mathbf{m}^{(c)}_{t,p}, \forall p\in\{1,2,\cdots, n\}$, where $\mathbf{h}_{t,p}$ is the transductive item embedding.
The embedding sequence without the last item $[\mathbf{z}^{(c)}_{t,1}, \cdots , \mathbf{z}^{(c)}_{t,(n-1)}]$ is passed to SASRec, which returns the corresponding predictive embeddings $[\mathbf{z}^{(c)\prime}_{t,1}, \cdots , \mathbf{z}^{(c)\prime}_{t,(n-1)}]$.
Then, {similar to Eq.~\eqref{eq:globalloss},} we use the binary cross-entropy loss with group-specific negative samples: 
% \begin{align}\label{eq:persloss}
%      \mathcal{L} =& 
%      \log{(\sigma((\mathbf{z}^{(c)}_{a})^{T}\mathbf{z}^{(c)\prime}_{k}))} + 
%      \log{(\sigma((\mathbf{z}^{(c)}_{b})^{T}\mathbf{z}^{(c)\prime}_{p}))} \nonumber \\
%      -& \log{(\sigma((\mathbf{z}^{(c)}_{p})^{T}\mathbf{z}^{(c)\prime}_{k}))} 
%      - \log{(\sigma((\mathbf{z}^{(c)}_{q})^{T}\mathbf{z}^{(c)\prime}_{p}))},
% \end{align}
\begin{equation}\label{eq:persloss}
    \mathcal{L}^{(pers)}_{t}=\sum_{p=1}^{n-1}\log{\sigma((\mathbf{z}^{(c)}_{a_{p}})^{T}\mathbf{z}^{(c)\prime}_{t,p})} -\sum^{n-1}_{p=1}\log{\sigma((\mathbf{z}^{(c)}_{t,(p+1)})^{T}\mathbf{z}^{(c)\prime}_{t,p})}
\end{equation}
where $\{i_{a_{p}} : \forall p \in \{1,2,\cdots,(n-1)\} \} \subset \mathcal{I}^{(c)}$ are negative samples from $\mathcal{I}^{(c)}$, {which is specific to the user group $\mathcal{U}^{(c)}$.}
Lastly, we update (1) the $\mathcal{U}^{(c)}$-specific PEFT module $\theta^{(c)}$ and projector $\texttt{MLP}^{(\theta)}$ and (2) the global components (transductive item embeddings and SASRec) via gradient descent to minimize $\mathcal{L}^{(pers)}_{t}$ (Eq.~\eqref{eq:persloss}).

\subsubsection{\textbf{Batch processing and training time}}\label{subsec:batchprocessing}
In practice, rather than updating with a single user as in the examples of Sections~\ref{subsec:globaltraining} and~\ref{subsec:ourdetail}, we form mini-batches with multiple users to accelerate training.
%A potential concern is that \method may require significantly more training time, since a single item could be re-encoded multiple times for different groups, each requiring its own multimodal embedding. 
%To address this, 
Especially, we construct each mini-batch from users within the same group, so items need to be encoded only {for that group} within the batch, as all users in the batch share the same item embeddings.
As a result, the training time of \method becomes comparable to that of Global PEFT, as detailed in Section~\ref{subsec:scalability}.

\subsection{Parameter overhead analysis}\label{subsec:paramoverhead}
In this section, we summarize the learnable components and their parameter counts.
In our experimental settings (see Section~\ref{subsec:setting}), when coupled with the best-performing PEFT method (spec., (IA)$^{3}$), 
\method with $8$ groups requires the following number of parameters for each component:  
(1) $\approx 368K$ for the PEFT modules, (2) $\approx 540K$ for the MLP projectors, (3) $470K-1M$ for the  transductive item embeddings, and (4) $\approx 13K$ for the SASRec backbone.
Considering that the foundation model (spec., CLIP) has over 151M parameters, the total number of learnable parameters in \method is relatively small, which is roughly 1.3\% of CLIP’s total.

% Due to this moderate model size, \method uses, on average, 12 GB of GPU memory per batch during training (see Appendix~\ref{subsec:parametercount}).
% Since \method assigns a PEFT module and MLP projector to each group, these components scale linearly with the number of groups.
% Nevertheless, in practice, this parameter increase is often not burdensome, as detailed in Appendix~\ref{subsec:parametercount}. 
% Specifically, compared to Global PEFT, \method coupled with (IA)$^{3}$ (1) has approximately $800K$ more parameters, which is less than 1\% of CLIP’s parameter size and (2) incurs only $\approx 0.8$ GB of additional per-batch GPU memory on average across datasets during training.}

    \section{Experiment}
    \label{sec:experiment}

In this section, we analyze the effectiveness of \method in the multimodal recommendation tasks.
Specifically, we investigate the following five research questions:
\begin{itemize}[leftmargin=*]
    \item \textbf{RQ1 (Effectiveness).} How effective is \method for multimodal recommendation compared to the baseline methods?
    \item \textbf{RQ2 (Scalability).} {How long is the training time of \method, and how many parameters does \method additionally require?}
    \item \textbf{RQ3 (Case study).} How do the recommendations produced by \method differ from those of Global PEFT?
    \item \textbf{RQ4 (Design goal).} Does \method meet its design objective by focusing on different item aspects across user groups?
    \item \textbf{RQ5 (Ablation).} How does the performance of \method change when its individual key components are removed?
\end{itemize}

\subsection{Setting}
\label{subsec:setting}
In this section, we elaborate on our experimental settings, covering (1) datasets, (2) evaluation protocol, (3) baseline methods, and (4) backbone PEFT modules. 
Further details are in Appendix~\ref{app:experimentdetail}.

\subsubsection{\textbf{Datasets}}
We use public e-commerce benchmark datasets sourced from Amazon reviews, using the latest release (2023)~\cite{hou2024bridging}.
We consider four widely used domains: (1) Sports \& Outdoors, (2) Toys \& Games, (3) Beauty \& Personal Care, and (4) Arts, Crafts \& Sewing. 
The dataset statistics of each domain are presented in Table~\ref{table:datastat}.
For each item, we use the product title (text) and product image (vision) as the two modalities.
Further details regarding our dataset, including a detailed domain description and our data preprocessing, are provided in Appendix~\ref{app:datadetail}.

\subsubsection{\textbf{Evaluation protocol}}
%For method evaluation, 
We use the leave-one-out protocol, which is widely used in sequential recommendation research~\cite{kang2018self, sun2019bert4rec}.
Specifically, for each user, we hold out the most recent purchase for testing and the second most recent for validation.
The remainder of each user's purchase history serves as the training sequence. We train the model on all users' training sequences and tune hyperparameters using the validation items, with further tuning details in Appendix~\ref{subsec:hyperparametertuning}. 
After selecting the validation-best configuration, we evaluate on the test set: for each user, we feed the training sequence joined with the validation item, rank items as described in Section~\ref{subsec:globaltraining}, and compare the ranking with the held-out test item.

\begin{table}[t]
\centering
\caption{\textit{Dataset statistics.} 
`Avg. Seq. Len.' indicates the mean length of users' purchase-history sequences, and `Density' indicates the total number of user–item interactions divided by the product of the numbers of users and items.}
\label{table:datastat}
\resizebox{0.95\linewidth}{!}{
\begin{tabular}{l | rrrr}
\toprule
Domain & \# Users & \# Items & Avg. Seq. Len. & Density \\
\midrule
Sports \& Outdoors     & 25,363 & 15,701 & 7.5991 & 0.0005 \\
Toys \& Games          & 19,026 & 14,718 & 8.5330 & 0.0006 \\
Beauty \& Personal Care& 45,490 & 31,151 & 9.5379 & 0.0003 \\
Arts, Crafts \& Sewing & 24,511 & 18,884 & 9.4328 & 0.0005 \\
% Sports \& Outdoors     & 25,363 & 15,701 & 7.59910 & 0.00048 \\
% Toys \& Games          & 19,026 & 14,718 & 8.53295 & 0.00058 \\
% Beauty \& Personal Care& 45,490 & 31,151 & 9.53789 & 0.00031 \\
% Arts, Crafts \& Sewing & 24,511 & 18,884 & 9.43279 & 0.00050 \\
\bottomrule
\end{tabular}}
\end{table}

\subsubsection{\textbf{Baseline methods and \method}}\label{subsec:baselines}
For comparison, we evaluate seven sequential recommender system baselines spanning five method categories. 
{We additionally report comparisons with non-sequential baseline methods in Appendix~\ref{app:sotamm}.}
\begin{itemize}[leftmargin=*]
    \item \textbf{Type 1 (Non-multimodal method).} The first type is a method that does not use any multimodal information, solely relying on the user-item interactions.
    Specifically, we use naive SASRec~\cite{kang2018self}, our backbone sequential recommender system.
    We call this method \textbf{w/o MM}.
    
    \item \textbf{Type 2 (Frozen multimodal encoder).} The second type is a method that uses multimodal information but does not fine-tune the multimodal foundation model. 
    It shares the same architecture as Global PEFT, but keeps CLIP frozen (i.e., no PEFT updates). 
    We call this method \textbf{Frozen MM}.
    
    \item \textbf{Type 3 (Global PEFT).} The third type is Global PEFT, which fine-tunes the multimodal foundation model via PEFT, but all users share the same PEFT module, as detailed in Section~\ref{subsec:methodbaseline}.
    
    \item \textbf{Type 4 (Embedding-based user-level personalization).} This variant of Global PEFT incorporates learnable user embeddings—a standard personalization technique in sequential recommendation~\cite{wu2024personalized,wu2020sse}. 
    We use two approaches: (1) \emph{user-as-sequence-element}, where a user embedding is inserted into the last part of the input purchase sequence as an additional element~\cite{wu2024personalized}, and (2) \emph{user–item concatenation}, where the user embedding is concatenated with each item embedding before being fed to the backbone~\cite{wu2020sse}.
    We call the two methods \textbf{User-level 1} and \textbf{User-level 2}, respectively.

    \item \textbf{Type 5 (Embedding-based group-level personalization).}
    This variant of Global PEFT also employs additional embeddings, analogous to the Type 4 baselines. 
    However, instead of user-level embeddings, we use \emph{group-level} embeddings: each group is associated with an embedding, and all users assigned to that group share it. 
    As in Type 4, we consider two methods—(i) group-as-sequence-element and (ii) group–item concatenation—which we denote \textbf{Group-level 1} and \textbf{Group-level 2}, respectively.
\end{itemize}
For \method, we set the number of user groups to $8$ for all settings, with additional results for (1) other group counts and (2) group size distributions in Appendices~\ref{app:clustercount} and~\ref{app:clustersize}, respectively.
We provide further training details and hyperparameter configurations of baseline methods and \method in Appendix~\ref{subsec:hyperparametertuning}

\subsubsection{\textbf{Backbone PEFT}}
We use three backbone PEFT modules: LoRA~\cite{hu2021lora}, (IA)$^3$~\cite{liu2022fewshot}, and IISAN~\cite{fu2024iisan}. 
LoRA and (IA)$^3$ are general-purpose, Transformer-targeted methods widely used to adapt multimodal foundation models, whereas IISAN is tailored for multimodal recommendation.
For LoRA, following \citet{hu2021lora}, we insert low-rank adapters into the query and value projection matrices of the attention layers. 
For (IA)$^{3}$, following \citet{liu2022fewshot}, we apply gating functions to the key and value projections and the first linear layer of the feed-forward network. 
For IISAN, we adopt the original adapter architecture from \citet{fu2024iisan}.

% \subsubsection{\textbf{Training and hyperparameters}}\label{subsec:hyperparametertuning}
% All methods are trained with the AdamW optimizer~\cite{loshchilov2017decoupled}. 
% We train all the baseline methods for 30 epochs, evaluate validation performance after each epoch using the corresponding checkpoint, and select the checkpoint with the highest validation Hit Ratio@30.
% For \method, we first train Global PEFT for 10 epochs, then perform user grouping, and finally train \method for 20 epochs, selecting the best checkpoint from these 20 epochs.
% We fix the dimensions of the transductive item embeddings and SASRec's hidden layer at 32 for all models, and set the dropout rate to 0.3.
% We tune the learning rate over $\{10^{-4},\,5\times10^{-5},\,10^{-5}\}$ and the weight decay over $\{5\times10^{-4},\,10^{-4},\,5\times10^{-5},\,10^{-5}\}$.
% We set the number of user groups to $8$ for all settings, with additional results for other group counts in Appendix~\ref{subsec:grouping}.
% We set the LoRA rank and the IISAN's adapter hidden dimension as $4$ and $32$, respectively.

\begin{table*}[t]
\centering
\small
\setlength{\tabcolsep}{2.0pt} % default ~6pt
\caption{\textit{(RQ1) Multimodal recommendation performance}.
All metrics are multiplied by 100 for better readability. 
Numbers in parentheses indicate standard deviation. 
H@K and N@K denote Hit-Ratio@K and NDCG@K, respectively. 
Best results are highlighted with a \gframe{\textbf{green}} box. 
Notably, \method outperforms the baseline methods in 44 of 48 settings.}
\label{tab:performance}
{
\renewcommand{\arraystretch}{0.8}
\resizebox{\linewidth}{!}{
\begin{tabular}{c | l| cccc | cccc | cccc | cccc}
\toprule

\multirow{2}{*}{PEFT} & 
\multirow{2}{*}{Methods} & \multicolumn{4}{c|}{Sports \& Outdoors} & \multicolumn{4}{c|}{Toys \& Games} & \multicolumn{4}{c|}{Beauty \& Personal Care} & \multicolumn{4}{c}{Arts, Crafts \& Sewing} \\
&  & H@20 & H@30 & N@20 & N@30 & H@20 & H@30 & N@20 & N@30 & H@20 & H@30 & N@20 & N@30 & H@20 & H@30 & N@20 & N@30 \\

\midrule

\multirow{2}{*}{N/A} & w/o MM         
& 2.79 & 3.76 & 1.08 & 1.29  & 2.27  & 3.04  & 0.91  & 1.05  & 2.60  & 3.47  & 1.12  & 1.35  & 3.91  & 5.20  & 1.48  & 1.76  \\

& Frozen MM    
& 4.11  & 5.38  & 1.71  & 1.99  & 3.24  & 4.34  & 1.31  & 1.55  & 3.60  & 4.66  & 1.45  & 1.67  & 4.87  & 6.42  & 1.89  & 2.21  \\

\midrule 
\midrule 

\multirow{7}{*}{LoRA} & Global PEFT          
& 4.64  & 5.97  & 1.94  & 2.22  & 4.19  & 5.67  & 1.71  & 2.02  & 4.15  & 5.52  & 1.72  & \best 2.02  & 6.01  & 7.81  & 2.42  & 2.81  \\

\cmidrule{2-18} 

& User-level 1 % wu2024personalized
& 4.48  & 5.96  & 1.86  & 2.18  & 4.27  & 5.67  & 1.74  & 2.04  & 4.19  & 5.52  & 1.68  & 1.95  & 6.03  & 7.92  & 2.44  & 2.83  \\

& User-level 2 % wu2020sse
& 4.51  & 5.86  & 1.90  & 2.19  & 4.13  & 5.46  & 1.67  & 1.91  & 4.03  & 5.43  & 1.66  & 1.94  & 6.05  & 7.91  & 2.43  & 2.82  \\

& Group-level 1
& 4.44  & 5.89  & 1.87  & 2.18  & 4.24  & 5.64  & 1.76  & 2.07  & 4.13  & 5.45  & 1.65  & 1.94  & 5.98  & 7.87  & 2.44  & 2.82  \\

& Group-level 2
& 4.56  & 5.94  & 1.92  & 2.21  & 4.26  & 5.64  & 1.75  & 2.06  & 4.14  & 5.47  & 1.70  & 1.98  & 5.96  & 7.94  & 2.45  & 2.88  \\

\cmidrule{2-18} 
& \method (Ours)
& \best 4.82 \std{.04} & \best 6.24 \std{.03} & \best 2.05 \std{.02} & \best 2.36 \std{.02}  & \best 4.79 \std{.03} & \best 6.15 \std{.04} & \best 1.97 \std{.03} & \best 2.25 \std{.02} & \best 4.27 \std{.03} & \best 5.58 \std{.03} & \best 1.74 \std{.01} & 2.01 \std{.01} & \best 6.62 \std{.11} & \best 8.64 \std{.06} & \best 2.76 \std{.04} & \best 3.17 \std{.02} \\

\midrule 
\midrule 

\multirow{7}{*}{(IA)$^{3}$} & Global PEFT          
& 4.55  & 5.97  & 1.90  & 2.21  & 4.58  & 5.96  & 1.89  & 2.19  & 4.25  & 5.61  & 1.73  & 2.03  & 6.24  & 8.16  & 2.59  & 3.00  \\

\cmidrule{2-18} 

& User-level 1 % wu2024personalized
& 4.45  & 5.92  & 1.89  & 2.20  & 4.41  & 5.86  & 1.82  & 2.13  & 4.25  & 5.55  & 1.76  & 2.02  & 6.09  & 8.06  & 2.49  & 2.92  \\

& User-level 2 % wu2020sse
& 4.39  & 5.74  & 1.81  & 2.11  & 3.95  & 5.38  & 1.69  & 1.95  & 3.64  & 5.20  & 1.62  & 1.89  & 5.68  & 7.64  & 2.32  & 2.71 \\

& Group-level 1
& 4.51  & 5.95  & 1.91  & 2.21  & 4.37  & 5.75  & 1.80  & 2.11  & 4.16  & 5.47  & 1.71  & 1.99  & 6.07 & 7.98 & 2.49 & 2.92 \\

& Group-level 2
& 4.58 & 5.99 & 1.93 & 2.23 & 4.36 & 5.87 & 1.87 & 2.18 & 4.14 & 5.48 & 1.71 & 2.00 & 6.17 & 8.02 & 2.52 & 2.94 \\

\cmidrule{2-18} 
& \method (Ours)                 
& \best 4.82 \std{.03} & \best 6.30 \std{.02} & \best 2.01 \std{.02} & \best 2.32 \std{.01} & \best 4.81 \std{.03} & \best 6.20 \std{.04} & \best 2.01 \std{.02} & \best 2.30 \std{.02} & \best 4.33 \std{.04} & \best 5.70 \std{.05}  & \best 1.78 \std{.04} & \best 2.06 \std{.03} & \best 6.56 \std{.02} & \best 8.50 \std{.03} & \best 2.67 \std{.01} & \best 3.11 \std{.02} \\

\midrule 
\midrule 

\multirow{7}{*}{IISAN} & Global PEFT      
& 4.77 & 6.13 & \best 2.01 & 2.30 & 4.50 & 5.97 & 1.89 & 2.15 & \best 4.31 & \best 5.71 & 1.74 & 2.05 & 6.02 & 7.89 & 2.46 & 2.85 \\

\cmidrule{2-18} 

& User-level 1 % wu2024personalized  
& 4.41 & 5.84 & 1.83 & 2.16 & 3.75 & 4.97 & 1.54 & 1.84 & 3.81 & 5.06 & 1.52 & 1.78 & 5.65 & 7.63 & 2.29 & 2.71 \\

& User-level 2 % wu2024personalized
& 4.39 & 5.79 & 1.82 & 2.15 & 4.23 & 5.54 & 1.75 & 2.04 & 4.03 & 5.42 & 1.67 & 1.95 & 5.86 & 7.82 & 2.41 & 2.75 \\

& Group-level 1
& 4.43 & 5.72 & 1.84 & 2.12 & 4.27 & 5.49 & 1.75 & 2.03 & 4.17 & 5.39 & 1.68 & 1.95 & 5.74 & 7.46 & 2.32 & 2.67 \\

& Group-level 2
& 4.38 & 5.67 & 1.84 & 2.12 & 4.05 & 5.38 & 1.68 & 1.97 & 4.14 & 5.34 & 1.72 & 1.97 & 5.64 & 7.38 & 2.27 & 2.64 \\

\cmidrule{2-18} 
& \method (Ours)                 
& \best 4.81 \std{.02} & \best 6.22 \std{.04} & 2.00 \std{.03} & \best 2.31 \std{.02} & \best 4.70 \std{.06} & \best 6.10 \std{.04} & \best 1.95 \std{.04} & \best 2.24 \std{.04} & 4.30 \std{.02} & 5.69 \std{.04} & 1.77 \std{.02} \best & 2.07 \std{.03}  \best & \best 6.37 \std{.05} & \best 8.17 \std{.03} & \best 2.59 \std{.04} & \best 2.98 \std{.03} \\

\bottomrule
\end{tabular}
}
}
\end{table*}

\subsection{RQ1. Recommendation performance}\label{subsec:mainperformance}

In this section, we analyze the effectiveness of \method in multimodal recommendation, in comparison with the baseline methods.

\subsubsection{\textbf{Setup}}
For baselines, we use seven methods described in Section~\ref{subsec:baselines}.
For each method, we run three trials with different random initializations and report the mean performance.
For the evaluation metrics, we use (1) Hit-Ratio@$K$ and (2) Normalized-Discounted-Cumulative-Gain@K (NDCG@$K$) with $K\in\{20, 30\}$.

\subsubsection{\textbf{Result.}}
As shown in Table~\ref{tab:performance}, \method outperforms all the baseline methods in most of the settings, demonstrating its strong performance.
Two points stand out.
First, \method outperforms Global PEFT in 44 out of 48 settings, indicating that assigning distinct PEFT modules to user groups is beneficial for the accurate multimodal recommendations. 
Second, \method outperforms user-embedding- or group-embedding-based personalization in every setting, suggesting that personalization at the PEFT-module level is more effective than relying on specialized embeddings in our case.

\begin{figure}[t] % t=top, b=bottom, h=here, ! to be less strict
  \centering
  \includegraphics[width=0.8\linewidth]{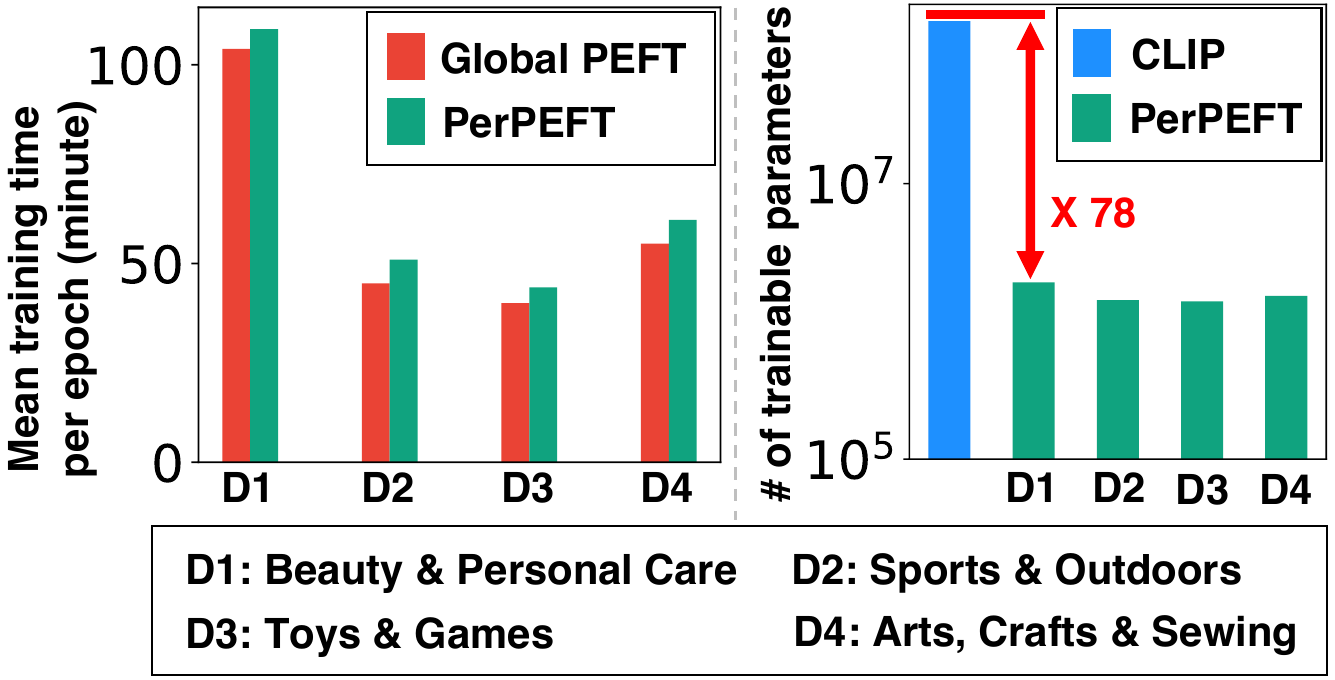} % or .png/.jpg
  \caption{\textit{(RQ2) Scalability analysis of \method}.
  Even with additional PEFT modules, \method incurs only a slight increase in training time relative to Global PEFT. 
  Moreover, the parameters introduced by \method are only 1.3\% of those of the multimodal foundation model (CLIP~\cite{radford2021learning}).}
  \label{fig:trainingtime}
\end{figure}

\subsection{RQ2. Scalability analysis}\label{subsec:scalability}

In this section, we conduct a scalability analysis of \method.

\subsubsection{\textbf{Setup.}}
We perform two analyses using (IA)$^{3}$ as the backbone PEFT method, which achieved the strongest performance among the three candidates. 
First, we investigate whether the additional PEFT modules in \method significantly affect training time relative to Global PEFT by comparing the average per-epoch training time across datasets (inference-speed analysis is provided in Appendix~\ref{subsec:inferencespeed}). 
Second, we compare the number of trainable parameters in \method with those of the multimodal foundation model, CLIP.

\subsubsection{\textbf{Result.}}
The results are in Figure~\ref{fig:trainingtime}. 
First, the incremental training time of \method over Global PEFT is small, indicating that the added PEFT modules do not impose a substantial training-time cost in practice. % —approximately 5–6 minutes per epoch across all datasets—
Second, even with personalization, the parameter increase remains minimal: the PEFT modules introduce only 1.3\% as many parameters as CLIP.

\begin{figure*}[t]
  \centering
  \begin{minipage}{\linewidth}
    \centering
    \includegraphics[width=0.97\linewidth]{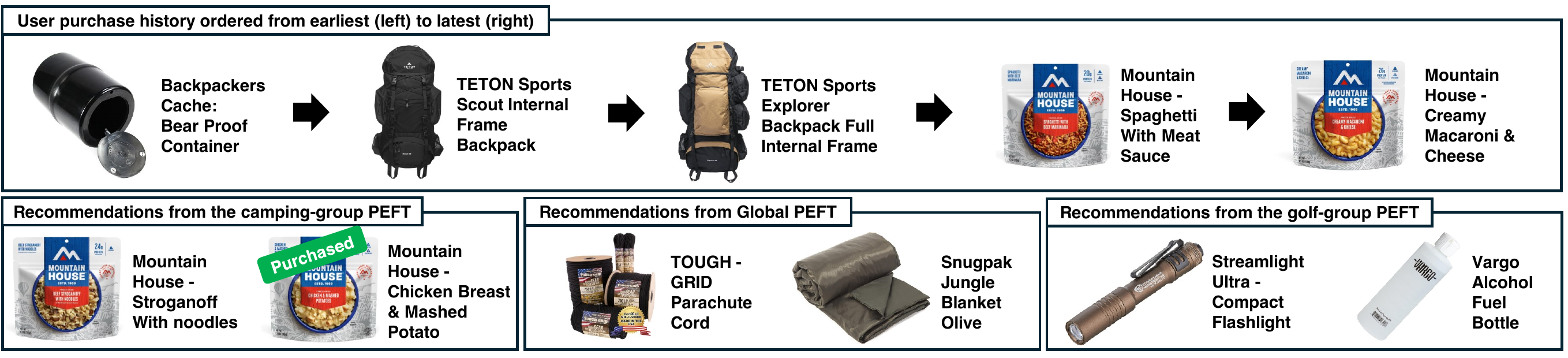}
    \vspace{-1mm}
    \captionof{subfigure}{\textit{Results in the Sports \& Outdoors dataset.}
    For a user whose recent interest is camping food, the camping-group PEFT accurately captures this preference and recommends relevant food items. 
    In contrast, Global PEFT and the golf-group PEFT recommend camping-related items but fail to identify camping foods specifically.}
    \label{fig:a}
  \end{minipage}\vspace{1mm}
  \begin{minipage}{\linewidth}
    \centering
    \includegraphics[width=0.97\linewidth]{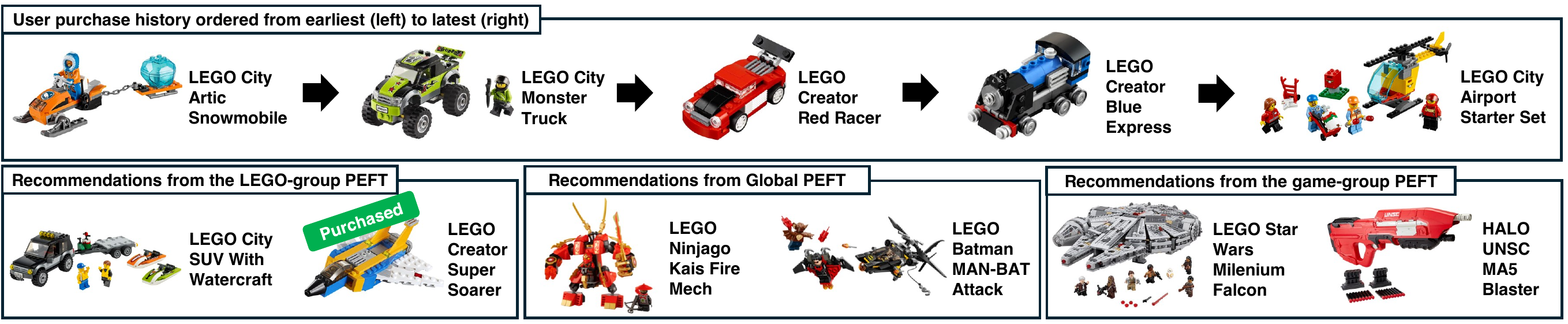}
    \vspace{-1mm}
    \captionof{subfigure}{\textit{Resuts in the Toys \& Games dataset.}
    For a user interested in vehicle-related LEGO items, the LEGO-group PEFT accurately captures this preference and recommends relevant vehicle-themed LEGO products. 
    However, Global PEFT and the game-group PEFT include at least one LEGO item but fail to identify the vehicle-related ones.}
    \label{fig:b}
  \end{minipage}
  \caption{\textit{(RQ3) Case study of recommendation results.} \textcolor{ForestGreen}{Green} box entitled `Purchased' indicates that the corresponding item is the ground-truth test item for the corresponding user. The group-specialized PEFT module in \method can capture fine-grained user interests within the target group. 
  In contrast, Global PEFT and PEFT modules specialized for other groups tend to capture only coarse, high-level user interests within the target group.}
  \label{fig:casestudy}
\end{figure*}

\subsection{RQ3. Case study}\label{subsec:casestudy}

In this section, we provide an in-depth analysis regarding how the recommendations produced by \method differ from those of several baselines through a case study.

\subsubsection{\textbf{Setup}}
From the Sports \& Outdoors, we first perform user grouping (Section~\ref{subsec:grouping}). 
Among the resulting user groups, we investigate two user groups: golf-enthusiasts and camping-enthusiasts, which we call the golf group and camping group, respectively.\footnote{In the golf group, 70\% of users purchased at least one item whose title contains the word `golf', whereas only 9\% of users did so in the overall dataset.
Similarly, in the camping group, 83\% of users purchased at least one item whose title contains the word `camp', while 41\% of users did so in the overall dataset.}
For a camping-enthusiast user, we compare recommendations from (1) Global PEFT, (2) \method's PEFT module assigned to the golf group (golf-group PEFT), and (3) \method's PEFT module assigned to the camping group (camping-group PEFT).
We apply a similar analysis for the Toys \& Games dataset, where we analyze LEGO-enthusiasts (LEGO group) and game-enthusiasts (game group).\footnote{In the LEGO group, 52\% of users purchased at least one item whose title contains the word `LEGO', whereas only 21\% of users did so in the overall dataset.
Similarly, in the game group, 93\% of users purchased at least one item whose title contains the word `game', while 46\% of users did so in the overall dataset.}
For a LEGO-enthusiast user, we compare recommendations from (1) Global PEFT, (2) game-group PEFT, and (3) LEGO-group PEFT.
We use LoRA as the PEFT module, and for each model, we report the two items with the highest recommendation scores.

\subsubsection{\textbf{Result}}
Our case study results are shown in Figure~\ref{fig:casestudy}.
In the Sports \& Outdoors dataset, the most recent interest of the camping-enthusiast user is camping food (Figure~\ref{fig:casestudy}(a)). 
The camping-group PEFT successfully captures this preference and recommends relevant food items, whereas Global PEFT and the golf-group PEFT recommend camping-related items but fail to pinpoint camping foods.
A similar pattern appears in the Toys \& Games dataset: the LEGO-enthusiast user is particularly interested in vehicle-related LEGO items (Figure~\ref{fig:casestudy}(b)). 
The LEGO-group PEFT identifies this fine-grained interest and recommends appropriate items, while Global PEFT and the game-group PEFT suggest at least one LEGO item, but not the vehicle-related ones.
These results suggest that the group-specialized PEFT module in \method can capture fine-grained user interests within the target group, whereas Global PEFT and modules specialized for other groups tend to capture only coarse, high-level interests.

\begin{table*}[t]
\centering
\small
\setlength{\tabcolsep}{2.5pt} % default ~6pt
\caption{\textit{(RQ5) Ablation study.}
All metrics are multiplied by 100 for better readability. 
Numbers in parentheses indicate standard deviation. 
H@K and N@K denote Hit-Ratio@K and NDCG@K, respectively. 
Best results are highlighted with a \gframe{\textbf{green}} box. 
Notably, \method outperforms its variants in 15 out of 16 settings.}
\label{tab:ablation}

{\renewcommand{\arraystretch}{0.9}
\resizebox{\linewidth}{!}{
\begin{tabular}{l| cccc | cccc | cccc | cccc}
\toprule

\multirow{2}{*}{Methods} & \multicolumn{4}{c|}{Sports \& Outdoors} & \multicolumn{4}{c|}{Toys \& Games} & \multicolumn{4}{c|}{Beauty \& Personal Care} & \multicolumn{4}{c}{Arts, Crafts \& Sewing} \\
& H@20 & H@30 & N@20 & N@30 & H@20 & H@30 & N@20 & N@30 & H@20 & H@30 & N@20 & N@30 & H@20 & H@30 & N@20 & N@30 \\

\midrule

w/o textual modality
& 3.87 & 5.15 & 1.62 & 1.86 & 3.85 & 4.92 & 1.58 & 1.81 & 4.00 & 5.29 & 1.61 & 1.88 & 4.95 & 6.55 & 2.01 & 2.35 \\

w/o visual modality
& 4.80 & 6.16 & 1.99 & 2.28 & 4.48 & 5.97 & 1.81 & 2.12 & 4.19 & 5.43 & 1.68 & 1.97 & \best 6.68 & 8.57 & 2.74 & 3.15 \\

w/o group-specific negatives
& 4.31 & 5.73 & 1.84 & 2.14 & 4.30 & 5.67 & 1.77 & 2.06 & 3.91 & 5.16 & 1.61 & 1.87 & 6.19 & 8.32 & 2.50 & 2.96 \\

Large Global PEFT
& 4.51 & 5.95 & 1.86 & 2.17 & 4.00 & 5.42 & 1.67 & 1.94 & 4.01 & 5.32 & 1.65 & 1.89 & 5.94 & 7.72 & 2.41 & 2.78 \\

Random grouping
& 4.09 & 5.46 & 1.71 & 2.00 & 3.65 & 4.84 & 1.48 & 1.73 & 3.82 & 4.95 & 1.50 & 1.74 & 5.43 & 7.18 & 2.17 & 2.54 \\

\midrule

\method (Ours)                 
& \best 4.82 \std{.04} & \best 6.24 \std{.03} & \best 2.05 \std{.02} & \best 2.36 \std{.02}  & \best 4.79 \std{.03} & \best 6.15 \std{.04} & \best 1.97 \std{.03} & \best 2.25 \std{.02} & \best 4.27 \std{.03} & \best 5.58 \std{.03} & \best 1.74 \std{.01} & \best 2.01 \std{.01} & 6.62 \std{.11} & \best 8.64 \std{.06} & \best 2.76 \std{.04} & \best 3.17 \std{.02} \\

\bottomrule
\end{tabular}
}
}

\end{table*}

\begin{figure}[t] 
  \centering
  \includegraphics[width=1.0\linewidth]{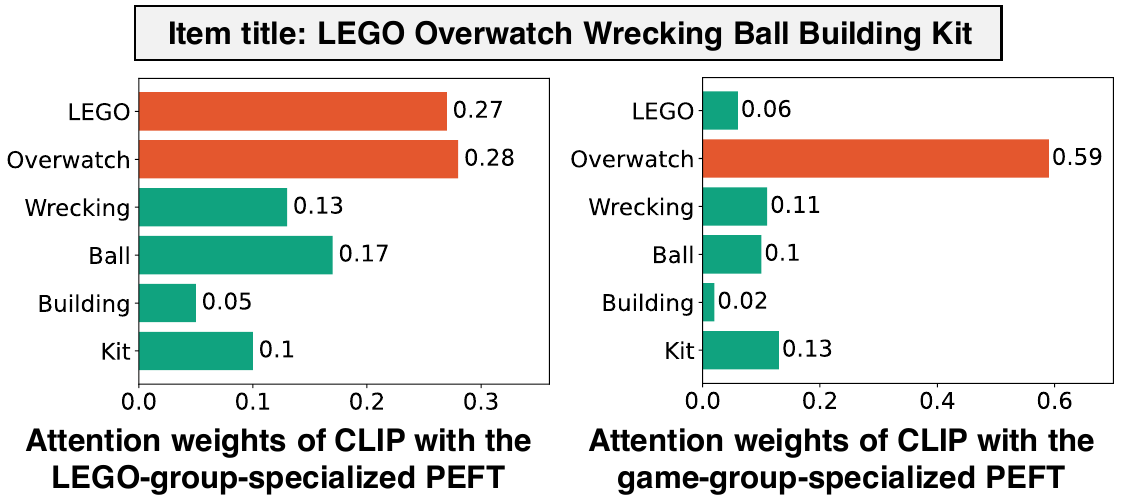} % or .png/.jpg
  \caption{\textit{(RQ4) Achievement of \method's design objective.} PEFT modules specialized for different groups guide CLIP~\cite{radford2021learning}, a multimodal foundation model, to focus on different aspects of the same item.}
  \label{fig:design}
\end{figure}

\subsection{RQ4. Design objective analysis}\label{subsec:designobjective}

In this section, we conduct a case study to examine whether the design objective of \method, enabling multimodal foundation models to focus on aspects that vary with user interests, is achieved. 
We present a text-based analysis in this section and provide an image-based analysis in Appendix~\ref{subsec:imageanalysis}.
{Moreover, we provide a quantitative analysis over all items to assess whether our design objective is achieved in Appendix~\ref{app:quantitativedesign}.}

\subsubsection{\textbf{Setup}}
We use the models introduced in Section~\ref{subsec:casestudy}: golf-group PEFT and camping-group PEFT from the Sports \& Outdoors dataset, and LEGO-group PEFT and game-group PEFT from the Toys \& Games dataset. 
For the Sports \& Outdoors dataset, we visualize CLIP's last-layer self-attention from the \texttt{[EOS]} token to each item-title token, comparing CLIP with the golf-group PEFT versus CLIP with the camping-group PEFT.
We use a similar approach to the Toys \& Games dataset with the LEGO-group PEFT and the {game-group PEFT.}
We use LoRA as our backbone PEFT.

% From the Sports \& Outdoors dataset, we first perform user grouping (Section~\ref{subsec:grouping}). 
% Among the resulting user groups, we investigate two user groups: golf-enthusiasts and camping-enthusiasts, which we call the golf group and camping group, respectively.\footnote{In the golf group, 70\% of users purchased at least one item whose title contains the word `golf', whereas only 9\% of users did so in the overall dataset.
% Similarly, in the camping group, 83\% of users purchased at least one item whose title contains the word `camp', while 41\% of users did so in the overall dataset.}
% For the Sports \& Outdoors dataset, we visualize the last-layer attention weights of CLIP for a given item title, where each token's weight is aggregated into the \texttt{[CLS]} token—the final text embedding—when CLIP is attached with either (1) the PEFT assigned to golf-enthusiasts (golf-group PEFT) or (2) the PEFT assigned to camping enthusiasts (camping-group PEFT).
% For the Toys \& Games dataset, we analyze LEGO-enthusiasts (LEGO group) and game-enthusiasts (game group).\footnote{In the LEGO group, 52\% of users purchased at least one item whose title contains the word `LEGO', whereas only 21\% of users did so in the overall dataset.
% Similarly, in the game group, 93\% of users purchased at least one item whose title contains the word `game', while 46\% of users did so in the overall dataset.}
% We use a similar approach as in the Sports \& Outdoors dataset by using the LEGO-group PEFT and the camping-group PEFT.
% We use LoRA as our backbone PEFT method.

\subsubsection{\textbf{Result}}
As shown in Figures~\ref{fig:mainattention} and \ref{fig:design}, the attention patterns of CLIP vary substantially when equipped with PEFT modules specialized for different groups. 
Given an item title describing a black golf bag, the golf-group PEFT places strong emphasis on the token `Golf', whereas the camping-group PEFT focuses on `Travel' and `Bag', terms more closely associated with camping (Figure~\ref{fig:mainattention}). 
Likewise, for a LEGO character from the game Overwatch, the LEGO-group PEFT focuses on both `LEGO' and `Overwatch', while the game-group PEFT concentrates primarily on the token `Overwatch' (Figure~\ref{fig:design}). 
These results indicate that different PEFT modules guide CLIP to attend to distinct aspects of items depending on the user group for which the module is specialized, thereby supporting that \method achieves its design objective.

% \begin{figure*}[t]
%   \centering
%   \begin{minipage}{0.48\linewidth}
%     \centering
%     \includegraphics[width=\linewidth]{Figures/AttentionWeight1.pdf}
%     \captionof{subfigure}{\textit{Results in the Sports \& Outdoors dataset.} While CLIP with the golf-group PEFT focuses on the term `Golf', CLIP with the camping-group PEFT attends instead to `Travel' and `Bag'.}
%     \label{fig:a2}
%   \end{minipage}\hspace{3mm}
%   \begin{minipage}{0.48\linewidth}
%     \centering
%     \includegraphics[width=\linewidth]{Figures/AttentionWeight2.pdf}
%     \captionof{subfigure}{\textit{Resuts in the Toys \& Games dataset.}
%     While CLIP with the LEGO-group PEFT focuses both on `LEGO' and `Overwatch', CLIP with the game-group PEFT focuses only on `Overwatch'.}
%     \label{fig:b2}
%   \end{minipage}
%   \caption{\textit{(RQ4) Achievement of \method's design objective.} PEFT modules specialized for different groups guide CLIP~\cite{radford2021learning}, a multimodal foundation model, to focus on different aspects of the same item.}
%   \label{fig:design}
% \end{figure*}

\subsection{RQ5. Ablation study}\label{subsec:ablation}

In this section, we investigate whether each key component of \method is beneficial for recommendation performance. 

% We use five variants of \method, which are equipped with LoRA~\cite{hu2021lora} as the backbone PEFT method: 
\vspace{-1mm}
\subsubsection{\textbf{Setup}}
We use five variants, which use LoRA for PEFT: 
\begin{enumerate}[label=\textbf{V\arabic*}, leftmargin=*]
    \item (w/o textual modality). Not using textual information.
    \item (w/o visual modality). Not using visual information
    \item (w/o group-specific negatives). Not using the group-specific negative sampling described in Section~\ref{subsec:remedy}.
    \item (Large Global PEFT). Using a single large PEFT module instead of multiple group-specific modules. {The total number of parameters in Large Global PEFT is approximately the same as that in \method (within $\pm 10^{3}$ parameters).}
    \item (Random grouping). Randomly grouping users rather than considering their interests as in Section~\ref{subsec:grouping}.
\end{enumerate}
% (V1) not using the textual information (w/o textual modality), (V2) not using the visual information (w/o visual modality), (V3) not using the group-specific negative sampling strategy described in Section~\ref{subsec:remedy} (w/o group-specific negatives), (V4) using a single large PEFT module instead of multiple group-specific modules (Large Global PEFT)~\footnote{The total parameters in Large Global PEFT is approximately the same as \method (within $\pm 10^{3}$ parameters).} and (V5) randomly grouping users rather than grouping based on their interests as in Section~\ref{subsec:grouping} (random grouping).
% We employ the four evaluation metrics described in Section~\ref{subsec:mainperformance}.

\subsubsection{\textbf{Result}}
As shown in Table~\ref{tab:ablation}, all four variants underperform \method in most settings. 
These results indicate that (1) both text and visual modalities are necessary for strong performance, (2) grouping users by similar interests is important, 
(3) simply increasing the number of parameters in PEFT modules is not beneficial, and (4) group-specific negative sampling provides a significant performance gain.
Together, these results demonstrate that \method's core components are crucial for achieving strong performance.
\vspace{-1mm}

    \section{Conclusion}
    \label{sec:conclusion}
    In this work, we investigate personalized PEFT for multimodal recommendation. 
First, we introduce \method, a PEFT strategy that employs different PEFT modules for user groups with different interests, encouraging the multimodal foundation model to focus on item aspects that are beneficial for recommending items to the corresponding group. 
Second, we propose a specialized training strategy that uses group-specific negative sampling to help multimodal foundation models capture more fine-grained information that is beneficial for recommendations to the corresponding group. 
Third, through extensive experiments, we demonstrate that \method outperforms existing baseline methods in multimodal recommendation. 
%Code and datasets are in~\url{https://github.com/kswoo97/PerPEFT}.
% , {without substantially increasing training cost or parameter count.}
% Lastly, it is consistently effective when combined with various PEFT methods.
% {We provide our code and datasets~\cite{anongithub}.}
%, and (3) it remains lightweight, with PEFT parameters totaling only 1.6\% of those of the multimodal foundation model (CLIP).

{\small 
    \smallsection{Acknowledgements}
    This work was partly supported by the National Research Foundation of Korea (NRF) grant funded by the Korea government (MSIT) (No. RS-2024-00406985, 30\%).
    This work was partly supported by Institute of Information \& Communications Technology Planning \& Evaluation (IITP) grant funded by the Korea government (MSIT) (No. RS-2022-II220871, Development of AI Autonomy and Knowledge Enhancement for AI Agent Collaboration, 40\%)
    (No. RS-2024-00457882, AI Research Hub Project, 20\%)
    (No. RS-2019-II190075, Artificial Intelligence Graduate School Program (KAIST), 10\%).}

   % \newpage % please remove newpage
    
    \bibliographystyle{ACM-Reference-Format}
    \balance
	\bibliography{000Ref}
    
    \appendix

    \vspace{5mm}

    {\LARGE\noindent\textbf{Appendix}}

    \section{Details of parameter-efficient fine-tuning}
    \label{app:peftdetail}
    
In this section, we provide a detailed description of the parameter-efficient fine-tuning PEFT methods used in our work, which are LoRA~\cite{hu2021lora}, (IA)$^{3}$~\cite{liu2022fewshot}, and IISAN~\cite{fu2024iisan}.
\begin{itemize}
    \item \textbf{LoRA}~\cite{hu2021lora} is a method that learns the low-rank adaptation of the foundation models for the target downstream task.
    Specifically, consider a pre-trained parameter $\mathbf{W}\in \mathbb{R}^{d\times d'}$ of the foundation model.
    For the update of $\mathbb{W}$, we use two learnable matrices: $\mathbf{A}\in \mathbb{R}^{d\times k}$ and $\mathbf{B} \in \mathbb{R}^{k \times d'}$. 
    Commonly, $k$ is much smaller than $d$ and $d'$, and therefore, $\mathbf{A}$ and $\mathbf{B}$ are low-rank adapter for $\mathbf{W}$.
    The original $\mathbf{W}$ is replaced with $\mathbf{W}+\mathbf{A}\mathbf{B}$, and the $\mathbf{A}\mathbf{B}$ is updated to learn the adaptation of $\mathbf{W}$, while $\mathbf{W}$ itself remains frozen.
    Thus, in \method, each user group is equipped with its own dedicated low-rank adapters.
    
    \item \textbf{(IA)$^{3}$}~\cite{liu2022fewshot} is a method that rescales the dimension of intermediate embeddings to adapt the foundation model for the target downstream task.
    Specifically, consider an intermediate embedding $\mathbf{z} \in \mathbb{R}^{d}$ of a data point within the foundation model.
    Here, we use a learnable scalar vector $\mathbf{l} \in \mathbb{R}^{d}$ that rescales the values of $\mathbf{z}$.
    Thus, $\mathbf{z} \odot \mathbf{l}$ is used instead of $\mathbf{z}$, where $\odot$ is an element-wise product operation.
    Thus, in \method, each user group is equipped with its own dedicated scaling vectors.
    
    \item \textbf{IISAN}~\cite{fu2024iisan}, a PEFT method specially designed for multimodal recommendation, employs side-network-based adaptation, where gradients do not flow through the multimodal foundation model. 
    Instead, intermediate embeddings from the foundation model are injected into the side network as features at intermediate layers.
    In this process, gradients flow only through the side network, so the intermediate embeddings from the multimodal foundation model act as fixed vectors that can be cached. 
    Thus, in \method, each user group is equipped with its own dedicated side networks.
\end{itemize}

    \section{Details of dataset}
    \label{app:datadetail}
    
\subsection{Each domain description}

In this section, we describe each domain of the latest Amazon review dataset~\cite {hou2024bridging}.
\textbf{The Sports \& Outdoors} domain encompasses items such as sports equipment, camping gear, cycling tools, and other outdoor goods.
Further details can be found at \url{https://www.amazon.com/sports-outdoors}.
\textbf{The Toys \& Games} domain encompasses items such as dolls, puzzles, CD games, console games, and remote vehicles. Further details can be found at ~\url{https://www.amazon.com/toys}.
\textbf{The Beauty \& Personal Care} domain encompasses items such as hair care tools, skin care items, and makeup tools. 
Further details can be found at~\url{https://www.amazon.com/Beauty-Makeup-Skin-Hair-Products}.
\textbf{The Arts, Crafts \& Sewing} domain encompasses items such as fabric decorating items, drawing supplies, craft kits, and knitting goods.
Further details can be found at ~\url{https://www.amazon.com/Arts-Crafts-Sewing}.

\subsection{Pre-processing protocol}
We download the full dataset for each domain and preprocess the corresponding sequence data for our experiments. 
Because the original data contains nearly a million users, we subsample a fixed number of users to keep computation within our time budget.
Specifically, we subsample 20\% of users from domains exceeding one million users (Sports \& Outdoors; Toys \& Games; Beauty \& Personal Care) and 40\% from domains below one million users (Arts, Crafts, \& Sewing).
Afterward, following representative sequential recommendation work~\cite{kang2018self,sun2019bert4rec}, we retain users with at least five purchases and items with at least five purchases.
Then, we use the resulting datasets for recommendations, where the resulting datasets' statistics are provided in Table~\ref{table:datastat}.

    \section{Experimental details and additional results}
    \label{app:experimentdetail}
    
\begin{table}[!t]
\centering
\small
\setlength{\tabcolsep}{3.0pt} % default ~6pt
\caption{\textit{Comparison against non-sequential multimodal recommenders.}
H@20 and N@20 denote Hit-Ratio@20 and NDCG@20, respectively. 
Best results are highlighted with a \gframe{\textbf{green}} box. 
Notably, \method outperforms all the baselines.}
\label{tab:sotamm}

{\renewcommand{\arraystretch}{1.0}
\resizebox{\linewidth}{!}{
\begin{tabular}{l| cc | cc | cc }
\toprule

\multirow{2}{*}{Methods} & \multicolumn{2}{c|}{Sports \& Outdoors} & \multicolumn{2}{c|}{Toys \& Games} & \multicolumn{2}{c}{Arts, Crafts, \& Sewing} \\

& H@20 & N@20 & H@20 & N@20 & H@20 & N@20  \\

\midrule

LATTICE~\cite{zhang2021mining} & 4.69 & 1.98 & 4.06 & 1.66 & 6.09 & 2.44 \\

BM3~\cite{zhou2023bootstrap} & 4.61 & 1.94 & 3.91 & 1.52 & 6.12 & 2.49 \\

\midrule

\method (Ours) & 4.82 \best & 2.01 \best & 4.81 \best & 2.01 \best & 6.56 \best & 2.67 \best \\

\bottomrule

\end{tabular}
}
}
\vspace{3mm}
\end{table}

\begin{table}[t]
\centering
\small
\setlength{\tabcolsep}{3.0pt} % default ~6pt
\caption{\textit{Performance with LightGCN backbone.}
H@20 and N@20 denote Hit-Ratio@20 and NDCG@20, respectively. 
Best results are highlighted with a \gframe{\textbf{green}} box. 
Notably, \method outperforms GlobalPEFT baselines.}
\label{tab:lightgcn}

{\renewcommand{\arraystretch}{1.0}
\resizebox{\linewidth}{!}{
\begin{tabular}{l| cc | cc | cc }
\toprule

\multirow{2}{*}{Methods} & \multicolumn{2}{c|}{Sports \& Outdoors} & \multicolumn{2}{c|}{Toys \& Games} & \multicolumn{2}{c}{Arts, Crafts, \& Sewing} \\

& H@20 & N@20 & H@20 & N@20 & H@20 & N@20  \\

\midrule

GlobalPEFT & 4.34 & 1.81 & 3.32 & 1.32 & 5.00 & 1.97 \\

\midrule

\method (Ours) & 4.57 \best & 1.94 \best & 4.00 \best & 1.60 \best & 5.51 \best & 2.22 \best \\

\bottomrule

\end{tabular}
}
}
\end{table}

\subsection{Machines and implementation}

All experiments were conducted on (1) two Intel Xeon Silver 4214R CPUs and (2) eight NVIDIA RTX 8000 D6 (48GB) GPUs.
All deep learning methods were implemented with Pytorch 1.12.1.
% Further details regarding our implementation are provided in~\cite{anongithub}.

\subsection{\textbf{Training and hyperparameters}}\label{subsec:hyperparametertuning}
All methods are trained with the AdamW optimizer~\cite{loshchilov2017decoupled}. 
To avoid a memory explosion, following \citet{fu2024iisan}, we set a maximum sequence length of 10 per user, keeping the most recent items and dropping older ones.
We train all the baseline methods for 30 epochs, evaluate validation performance after each epoch using the corresponding checkpoint, and select the checkpoint with the highest validation Hit Ratio@30.
For \method, we first train Global PEFT for 10 epochs, then perform user grouping, and finally train \method for 20 epochs, selecting the best checkpoint from these 20 epochs.
We fix the dimensions of the transductive item embeddings and SASRec’s hidden layer at 32 for all models, and set the dropout rate to 0.3.
Moreover, for every method, we stack two self-attention blocks in SASRec with four attention heads, and fix the mini-batch size to $16$.
We tune the learning rate over $\{10^{-4},\,5\times10^{-5},\,10^{-5}\}$ and the weight decay over $\{5\times10^{-4},\,10^{-4},\,5\times10^{-5},\,10^{-5}\}$.
We set the LoRA rank and the IISAN's adapter hidden dimension as $4$ and $32$, respectively.
Note that (IA)$^{3}$ does not require hyperparameters.

\subsection{Non-sequential recommender analysis}

\subsubsection{\textbf{Comparing with non-sequential multimodal models}}\label{app:sotamm}

\noindent\textbf{Setup.} We compare \method coupled with (IA)$^{3}$ against two non-sequential multimodal recommenders: LATTICE~\cite{zhang2021mining} and BM3~\cite{zhou2023bootstrap}. 

\noindent\textbf{Results.} As shown in Table~\ref{tab:sotamm}, \method outperforms the two baselines, demonstrating its superiority over representative non-sequential recommender systems.

\subsubsection{\textbf{Performance under non-sequential models}}\label{app:lightgcn}
Recall that \method was primarily coupled with sequential recommender systems. 
We evaluate \method when coupled with a non-sequential recommender system, 

\noindent\textbf{Setup.} We use LightGCN~\cite{he2020lightgcn} as the backbone recommender system. 
As in the SASRec setting, we add item multimodal embeddings to the transductive item embeddings and then perform graph propagation. 
We use IISAN as our backbone PEFT module.

\noindent\textbf{Results.} As shown in Table~\ref{tab:lightgcn}, \method consistently outperforms GlobalPEFT, indicating that its effectiveness extends beyond sequential recommendation settings.

\begin{figure*}[t]
  \centering
  \begin{minipage}{0.23\linewidth}
    \centering
    \includegraphics[width=\linewidth]{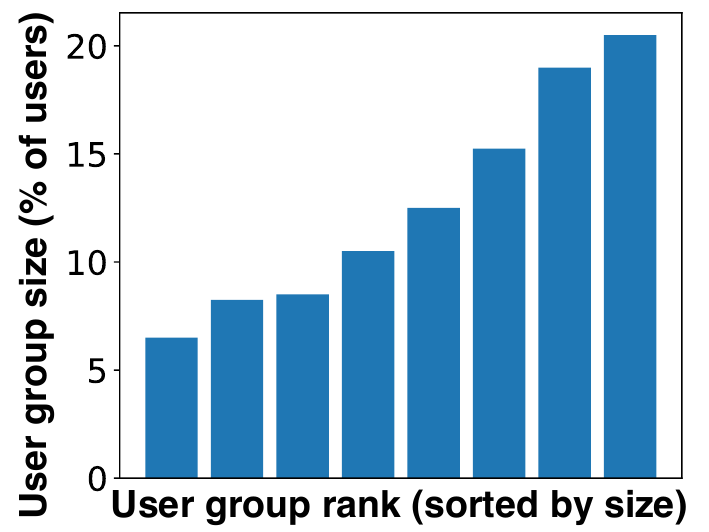}
    \vspace{-4mm}
    \captionof{subfigure}{Group size distribution.}
    \label{fig:cluster_b}
  \end{minipage}
  \hspace{2mm}
  \begin{minipage}{0.75\linewidth}
    \centering
    \includegraphics[width=\linewidth]{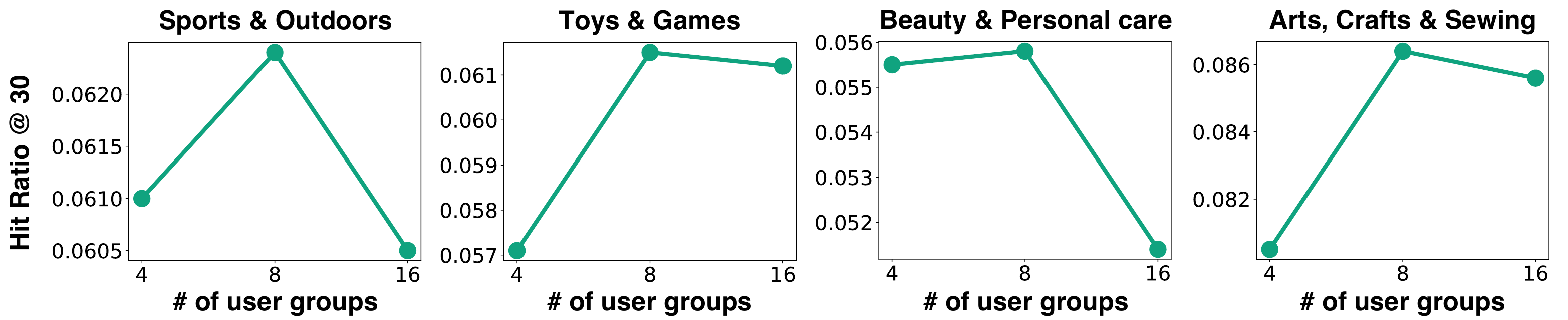}
    \vspace{-0.5mm}
    \captionof{subfigure}{Recommendation performance of \method under varying number of user groups.}
    \label{fig:cluster_a}
  \end{minipage}
  \caption{\textit{User group analysis.} User group sizes do not collapse (see (a)). In contrast, the performance of \method varies with the number of user groups, peaking when the number of user groups is 8 (see (b)).}
  \label{fig:cluster}
\end{figure*}

\begin{table}[t]
\centering
\small
\setlength{\tabcolsep}{2.0pt} % default ~6pt
\caption{\textit{Evaluation under various clustering algorithms.}
H@20 and N@20 denote Hit-Ratio@20 and NDCG@20, respectively. 
All numbers are multiplied by 100 for better readability.
Numbers in parentheses indicate standard deviations.
P-value reports the p-value from a paired t-test. 
Notably, the two clustering algorithms yield no statistically significant difference in performance.}
\label{tab:stattest}
{\renewcommand{\arraystretch}{1.0}
\resizebox{\linewidth}{!}{
\begin{tabular}{l| cc | cc | cc }
\toprule
\multirow{2}{*}{Methods} & \multicolumn{2}{c|}{Sports \& Outdoors} & \multicolumn{2}{c|}{Toys \& Games} & \multicolumn{2}{c}{Arts, Crafts, \& Sewing} \\
& H@20 & N@20 & H@20 & N@20 & H@20 & N@20  \\
\midrule
K-Means & 4.67 \std{0.09} & 1.96 \std{0.03} & 4.69 \std{0.07} & 1.94 \std{0.03} & 6.33 \std{0.07} & 2.60 \std{0.03} \\
GMM & 4.62 \std{0.03} & 1.95 \std{0.03} & 4.73 \std{0.04} & 1.94 \std{0.04} & 6.30 \std{0.03} & 2.62 \std{0.03} \\
\midrule 
P-value & 0.316 & 0.824 & 0.157 & 0.580 & 0.432 & 0.642 \\
\bottomrule
\end{tabular}
}
}
\end{table}

\begin{table}[t]
\centering
\small
\setlength{\tabcolsep}{3.0pt} % default ~6pt
\caption{\textit{Quantitative intra- and inter-cluster attention weights analysis.}
We measure the Jensen–Shannon divergence to compare attention weights. 
Notably, inter-cluster attention weight differences (Inter-clus.) are larger than intra-cluster differences (Intra-clus.).}
\label{tab:jsd}

{\renewcommand{\arraystretch}{1.0}
\resizebox{\linewidth}{!}{
\begin{tabular}{l| c | c | c }
\toprule

Attentions & Sports \& Outdoors & Toys \& Games & Arts, Crafts \& Sewing \\

\midrule

Intra-clus. & 0.0129 & 0.0109 & 0.0587 \\

Inter-clus. & 0.0637 & 0.0516 & 0.0778 \\

\bottomrule

\end{tabular}
}
}
\end{table}

\subsection{User group analysis}\label{subsec:clustersizeanalysis}

\subsubsection{\textbf{User group size analysis}}\label{app:clustersize}
We analyze the user group size distribution generated by K-Means.

\noindent{\textbf{Setup.}}
For each dataset, we sort groups in descending order of size. 
We then compute the mean group size for each rank across datasets.

\noindent{\textbf{Results.}}
As shown in Figure~\ref{fig:cluster}(a), even the smallest group contains 6.5\% of users, indicating that group-size collapse does not occur.

\subsubsection{\textbf{Number of user groups analysis}}\label{app:clustercount}
We evaluate recommendation performance under different numbers of user groups.

\noindent{\textbf{Setup.}}
We use three different user group numbers: $4$, $8$, and $16$.
We use LoRA~\cite{hu2021lora} as our backbone PEFT method.

\noindent{\textbf{Result.}}
As shown in Figure~\ref{fig:cluster}(b), \method with 8 groups outperforms the 4- and 16-group settings across all datasets. 
This suggests that both too few and too many groups are suboptimal.

\subsubsection{\textbf{Other clustering algorithm analysis}}\label{app:otherclustering}
Recall that \method was primarily coupled with the K-Means clustering algorithm.
We evaluate whether \method maintains strong performance under different clustering algorithms.

\noindent\textbf{Setup.}
We adopt a Gaussian Mixture Model (GMM) as the backbone clustering method and compare it against K-Means. 
We conduct 10 trials and perform a paired t-test across trials. 
We use IISAN as the backbone PEFT.

\noindent\textbf{Results.}
As shown in Table~\ref{tab:stattest}, the performance obtained with K-Means and GMM does not differ statistically significantly, indicating that \method is robust to the choice of clustering algorithm.

\subsection{Design objective analysis}\label{subsec:imageanalysis}

In this section, we further examine whether \method fulfills its design objective of personalized item-aspect concentration.

\subsubsection{\textbf{Quantitative analysis}}\label{app:quantitativedesign}
We provide quantitative evidence for the qualitative findings provided in Section~\ref{subsec:designobjective}.

\noindent\textbf{Setup.} We compute the Jensen–Shannon divergence (JSD) between the token-level attention distributions of the same item under (1) PEFT modules from different clusters and (2) PEFT modules from the same cluster trained with different random initializations (i.e., different seeds).
We repeat this analysis for all items and report the mean values.

\noindent\textbf{Results.} As shown in Table~\ref{tab:jsd}, inter-cluster JSD values are much higher than intra-cluster JSD values, demonstrating that our findings in Section~\ref{subsec:designobjective} are not specific to certain item cases.

\begin{figure*}[t]
  \centering
  \begin{minipage}{0.48\linewidth}
    \centering
    \includegraphics[width=\linewidth]{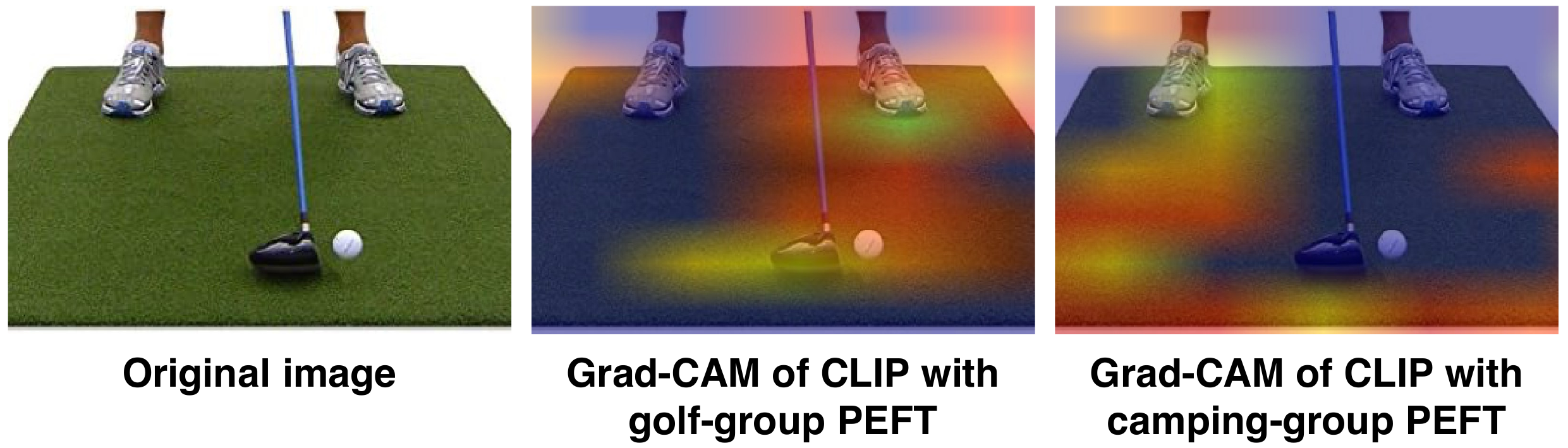}
    \captionof{subfigure}{\textit{Results in the Sports \& Outdoors dataset.} While CLIP with the golf-group PEFT focuses on the golf club and shoes, CLIP with the camping-group PEFT focuses on the green mat.}
    \label{fig:a3}
  \end{minipage}\hspace{3mm}
  \begin{minipage}{0.48\linewidth}
    \centering
    \includegraphics[width=\linewidth]{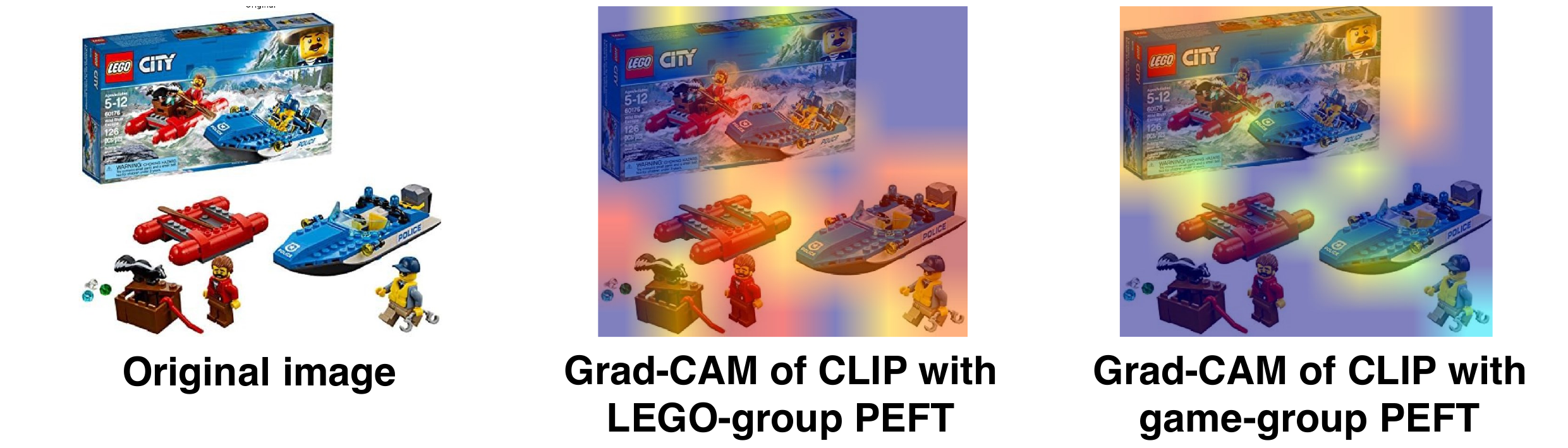}
    \captionof{subfigure}{\textit{Resuts in the Toys \& Games dataset.}
    While CLIP with the LEGO-group PEFT focuses both on characters and vehicles, CLIP with the game-group PEFT focuses on the LEGO box.}
    \label{fig:b3}
  \end{minipage}
  \caption{\textit{Achievement of \method's design objective.} PEFT modules specialized for different groups guide CLIP~\cite{radford2021learning}, a multimodal foundation model, to focus on different aspects of the same item.}
  \label{fig:design2}
\end{figure*}

\subsubsection{\textbf{Vision-level analysis}}
We analyze whether \method has achieved this design objective at the image level.

\noindent{\textbf{Setup}}
We follow the same protocol as in Section~\ref{subsec:designobjective}.
We use (1) an image of a man standing on a golf mat and holding a golf club from the Sports \& Outdoors dataset, and (2) an image of LEGO vehicles with the product packaging box from the Toys \& Games dataset.
Unlike in text-based analysis, we use GradCAM~\cite{selvaraju2017grad} to analyze CLIP behavior with the visual information.
Specifically, we visualize the focus of CLIP via GradCAM, and by this, we analyze which part CLIP focuses on.

\noindent{\textbf{Result}}
As shown in Figure~\ref{fig:design2}, the Grad-CAM results of CLIP vary substantially when equipped with PEFT modules specialized for different groups. 
Given an image of a man standing on the golf mat, the golf-group PEFT places strong emphasis on the token golf club and the man's shoes, whereas the camping-group PEFT focuses on the green mat, which is similar to the grass background widely seen in camping items (Figure~\ref{fig:design2} (a)). 
Likewise, for an image of LEGO characters and their packaging box, the LEGO-group PEFT focuses on LEGO characters and their vehicles, while the game-group PEFT concentrates primarily on the packaging box, which is widely seen in game products (Figure~\ref{fig:design2} (b)). 
These results indicate that different PEFT modules guide CLIP to attend to distinct aspects of items depending on the user group for which the module is specialized, thereby supporting that \method achieves its design objective.

    \section{Scalability analysis and limitations}
    \label{app:method}
    \begin{table}[t]
  \centering
  \small
  \setlength{\tabcolsep}{3.0pt} % default ~6pt
  \caption{\textit{Parameter counts of Global PEFT and \method.} (A) denotes the group-specific components and (B) denotes the global components.
  Compared to Global PEFT, \method requires approximately $800K$ parameters, which is less than $1$\% of the CLIP's parameter counts.}
  \label{tab:paramcount}
  \begin{tabular}{l | cccc}
    \toprule
    & Sports & Toys & Beauty & Arts \\
    \midrule
    \midrule
    (A) PEFT module (IA)$^3$ & \multicolumn{4}{c}{46,080} \\
    (A) MLP projector & \multicolumn{4}{c}{67,872} \\
    (B) SASRec & \multicolumn{4}{c}{12,992} \\
    \midrule 
    (B) \makecell[l]{Transductive \\embeddings} & 502,816 & 471,360 & 997,216 & 604,672 \\
    \midrule 
    Global PEFT & 629,760 & 598,304 & 1,124,160 & 731,616 \\
    \method & 1,427,424 & 1,395,968 & 1,921,824 & 1,529,280 \\
    \midrule
    CLIP~\cite{radford2021learning} & \multicolumn{4}{c}{151,323,393} \\
    \bottomrule
  \end{tabular}
\end{table}

\begin{table}[!t]
  \centering
  \small
  \setlength{\tabcolsep}{2.5pt} % default ~6pt
  \caption{\textit{Average per-batch GPU memory consumption during training of Global PEFT and \method (GB).} 
  \method takes only 0.8 GB more GPU memory than Global PEFT on average.}
  \label{tab:memorycount}
  \begin{tabular}{l | cccc | c }
    \toprule
    & \makecell[l]{Sports} & \makecell[l]{Toys} & \makecell[l]{Beauty} & \makecell[l]{Arts} & Average \\
    \midrule
    Global PEFT & 10.7 & 10.8 & 11.4 & 12.4 & 11.3\\
    \method & 11.2 & 11.3 & 13.3 & 12.7 & 12.1\\
    \bottomrule
  \end{tabular}
\end{table}

In this section, we provide scalability analysis and limitations regarding our proposed method, \method.

\subsection{Parameter overhead analysis}\label{subsec:parametercount}

In this section, we provide detailed results regarding our parameter overhead analysis presented in Section~\ref{subsec:paramoverhead}.

\subsubsection{\textbf{Setup.}} 
We present component-level parameter counts for \method and Global PEFT (Section~\ref{subsec:paramoverhead}) and the average per-batch GPU memory required to train each method across datasets.
We use (IA)$^{3}$~\cite{liu2022fewshot} as our backbone PEFT method.

\subsubsection{\textbf{Results.}}
As shown in Table~\ref{tab:paramcount}, \method’s additional parameters over Global PEFT are marginal—less than 1\% of the CLIP backbone’s total parameter count~\cite{radford2021learning}.
Moreover, owing to the small parameter increase, \method uses only 0.8 GB more GPU memory on average during training than Global PEFT (refer to Table~\ref{tab:memorycount}), indicating that the overhead is marginal in practice.

% \begin{table}[t]
%   \centering
%   \small
%   \setlength{\tabcolsep}{3.0pt} % default ~6pt
%   \caption{\textit{Per-user SASRec inference time (seconds).} 
%   On average, SASRec only takes 0.0082 seconds to produce a recommendation for a single user.}
%   \label{tab:inferencetime}
%   \begin{tabular}{cccc|c}
%     \toprule
%     \makecell[l]{Sports} & \makecell[l]{Toys} & \makecell[l]{Beauty} & \makecell[l]{Arts} & Average\\
%     \midrule
%     0.0073 & 0.0074 & 0.0098 & 0.0082 & 0.0082\\ 
%     \bottomrule
%   \end{tabular}
% \end{table}

\subsection{Inference time analysis}\label{subsec:inferencespeed}

We analyze the inference scalability of \method, comparing with that of Global PEFT.

\subsubsection{\textbf{Item embedding pre-computation}}
After training, item embeddings can be precomputed and cached for inference. 
Since each user group has its own PEFT module, we compute one set of item embeddings per group. Consequently, the offline precomputation time for \method\ scales linearly with the number of groups $C$ (e.g., $\approx 8 \times$ the cost of Global PEFT in our experimental setting). 
We would like to emphasize that this is a one-time, offline cost; embeddings need not be recomputed per user at inference time.

\subsubsection{\textbf{User inference}}

For users seen during training, \method runs at the same inference time as Global PEFT because group membership is known and group-specific item embeddings are cached. 
For unseen users, we (1) assign the user to a group and then (2) score items using that group’s embeddings. 
This extra assignment step adds additional time; in our implementation, inference for unseen users is approximately 2$\times$ that for seen users.
% However, it is important to note that the inference time for each user is marginal, which only takes 0.0082 seconds on average across datasets.

\subsection{Limitations of \method}

In this section, we discuss the potential limitations of \method.

\subsubsection{\textbf{Lack of theoretical analysis.}}
Although \method achieves stronger results than Global PEFT in the multimodal recommendation tasks, a principled theoretical understanding, such as provable effects on CLIP’s representational effectiveness, remains open. 
The joint nonlinear structure of the backbone and PEFT layers complicates analysis, pointing to an important direction for future work.

\subsubsection{\textbf{Lack of modality information}}
In practice, certain items can be modality-incomplete (e.g., missing text or images)~\cite{wu2024deep}. 
Since our method requires both textual and visual inputs, developing a version of \method that can be used under missing modalities would enhance its real-world applicability.

\subsubsection{\textbf{Lack of handling users with multiple interests}}
While we adopt hard user assignments to clusters based on user preferences, some users may exhibit interests that span multiple clusters. 
Consequently, developing techniques that can effectively model such users could further improve performance.

\end{document}